\documentclass[aps,preprint]{revtex4-1}

\usepackage{amssymb,amsfonts,amsmath,esint,tabularx}
\usepackage[breaklinks=true,urlcolor=blue,bookmarks=true]{hyperref}
\hypersetup{pdfauthor={Jiang Qian},pdfkeywords={Dielectric Enhancement, Zeta 
Potential, Electrolyte Solution, Universal Scaling},pdfsubject={Theoretical 
Physics, Classical Electrodynamics, Integral Equation},pdftitle={Dielectric 
Enhancement without zeta-Potential Exhibits Universal Scaling}}

\begin{document}

\title{Universal Dielectric Enhancement from Externally Induced Double Layer 
Without $\zeta$-potential}

\author{Jiang Qian and Pabitra N. Sen}
\affiliation{4097 Silsby Road, University Heights, OH 44118 and University of 
North Carolina, Chapel Hill, NC 27599}



\begin{abstract}
\textbf{Abstract:}
Motivated by recent experiments showing over $10^4$-fold increase in induced 
polarization from electrochemically inert, conducting materials in dilute 
saline solutions, we theoretically demonstrate a new mechanism for dielectric 
enhancement, in the absence of $\zeta-$potentials at interfaces between 
non-insulating particles and an electrolyte solution. We further show that the 
magnitude of such enhancement obeys universal scaling laws, independent of the 
particle's electrical properties and valid across particle shapes: for a dilute 
suspension of identical, but arbitrarily shaped particles of a linear dimension 
$a$ and volume fraction $f$, as $\omega\to0$ the effective real dielectric 
constant of the mixture is enhanced from that of water by a factor 
$1+f~(P_r+(a/\lambda)P_i)$, and the frequency-dependent phase shift of its 
impedance has a scale-invariant maximum  $f\,\mathsf{\Theta}$ if particles are 
much more conductive than the solution. Here $\lambda$ is the solution's Debye 
length and $P_r$, $P_i$, $\mathsf{\Theta}$ are dimensionless numbers determined 
solely by the particles' shape.  Even for a very dilute electrolyte solution 
(e.g.  $10^{-3}$ molar), sub-mm sized particles, at volume fraction $f=0.1$, 
can give a $10^4$-fold dielectric
enhancement, producing an easily observable  phase shift maximum in a simple 
impedance measurement.
We also derive frequency cutoffs as conditions for observing these 
enhancements, showing that insulating particles produce no enhancement without 
$\zeta$-potential.
To prove these results for particles of arbitrary shapes, we develop a physical 
picture where an externally induced double layer
(EIDL), in contrast to the Guoy-Chapman double layer on interfaces with 
significant $\zeta$-potentials, dominates the low-frequency dynamics and 
produces dielectric enhancement.

\textbf{Significance Statement:}
Dielectric enhancement is a striking and technologically important phenomenon 
in complex fluids: the introduction of a small amount of suspended particles 
can boost the low-frequency dielectric constant of an electrolyte solution by 
many orders of magnitudes. It is crucial for understanding dielectric responses 
in systems as disparate as gold particles in cytosol and mineral deposits in 
geological exploration.  Dielectric enhancement is traditionally thought to be 
caused by large surface potentials on solid-fluid interfaces. Our theory 
suggests that this phenomenon is more widespread and fundamental in nature.  We 
show that even without electrochemically active surfaces, a dilute suspension 
of non-insulating, arbitrarily shaped particles in an electrolyte solution can 
produce a huge low-frequency dielectric enhancement that obeys universal 
scaling laws.
\end{abstract}
\maketitle



In recent experiments aiming at designing contrast agents, intriguing 
preliminary data suggest that a dilute (volume fraction $f\le0.1$) suspension 
of sub mm-sized particles in a dilute electrolyte solution can produce real 
dielectric constant over $10^4$ times that of the water. These non-metallic 
conductive particles are not known to attract large surface charges and are 
chemically inert in the solution. These data are unexpected in the current 
theoretical framework for dielectric enhancement, underlying significant gaps 
in our understanding of this important phenomenon in a seemingly simple 
physical system.

At low-frequency ($\omega$) where the depth of penetration is large, the 
conductivity and  dielectric measurements have great potential as diagnostic 
probes into complex and hard to reach geometries like human bodies or porous 
media.  For example, both noble metal particles~\cite{gold_nanoparticle} and 
semiconductor quantum dots~\cite{quantum-dots}, functionalized with antibodies 
and selectively enriched in particular types of cells or organelles, have been 
used as contrast agents and biosensors \cite{Matijevic, Buck, Biosensors}.  
Beyond traditional optical probes, they may hold promise for noninvasive 
imaging applications like electrical impedance 
tomography~\cite{impedence-medical}.  As another example, in geological 
exploration,  shale oil and gas deposits have semiconducting pyrites inclusions 
and the latter can be detected by geophysical exploration techniques 
\cite{polarization-rock}. Artificially introduced contrast agents, mixed with 
saline fluids injected to wells, are being actively explored for uses in 
conductivity measurements.

Most theoretical explanations for dielectric enhancement to date rely on the 
Gouy-Chapman double layer created by a large $\zeta$-potential and surface 
charges on the solid-solution 
interface~\cite{Dukhin,Delacey,Obrien,Chew,Fixman,Hinch}.  This is natural 
because they largely focus on experiments with \emph{insulating} 
particles~\cite{Schwan_enhancement,Clay,Lyklema}, which we show below to be 
incapable of producing enhancement without $\zeta$-potentials.  But many 
contrast agents in medicine and engineering, not to mention natural mineral 
deposits, are semiconducting or conducting.
To our knowledge, only Wong~\cite{Wong} explicitly considered dielectric 
enhancement in the absence of $\zeta$-potential in the special case of fully 
conducting, ideally \emph{non}-polarizable metallic spheres. However, he 
included an amalgam of ion species both with and without Faradaic currents from 
redox reaction~\cite{Bard,Franschetti}, making his model needlessly complex and 
his results difficult to interpret.  Thus the simplest but crucial zeroth order 
``model organism'' of electrolyte solution-based complex fluid, where 
\emph{ideally polarizable} solid-fluid interfaces have \emph{no} 
$\zeta$-potential and \emph{no} redox reaction, has been curiously 
under-investigated in the literature.

Here we study such a model that strips the physics of dielectric enhancement to 
its bare essentials, with neither $\zeta$-potential nor Faradaic current. We 
show that a dilute suspension such particles, made from a non-insulating 
material, suffices to produce dielectric enhancement, the magnitude of which 
follows a universal scaling law whose parameters are determined solely by the 
\emph{shape} of the particles.  Due to the simplicity of our model, we are able 
to demonstrate this enhancement for particles of \emph{any} shapes.  We further 
derive conditions for observing enhancement and show that, for particles much 
more conductive than the solution, there is a significant, 
\emph{scale-invariant} maximum in the frequency-dependent phase shift in 
impedance measurements.
\section{Governing Equations and Boundary Conditions}
Consider electrically uniform solid bodies or immiscible fluid (henceforth 
simply referred to as ``solid'') submersed in an electrolyte solution as 
suspended particles or droplets. We assume the solid bodies are smooth and 
convex, and have aspect ratios not too far from one (\emph{i.e.} they are not 
needle- or disk-shaped). An external electric field $E_0\exp{(i\omega t)}$ in 
the $\widehat{z}$-axis drives the charge dynamics of the system.

We assume that the $\zeta$-potentials on all interfaces are
small.  In the online SI, we show that our analysis and results remain valid as 
long as the $\zeta$-potential $\psi_{\zeta}\ll k_B T/e$, such that the
it does not significantly alter the equilibrium ion densities distributions 
$N_{\pm}=N_0 e^{\mp e\psi_\zeta/k_B T}\approx N_0$. This condition can be 
satisfied by natural materials such as oxidized pyrite~\cite{pyrite}, as well 
as by materials whose $\zeta$-potential can be tuned by the pH of the 
solution~\cite{gold_zeta}.

To emphasize physics and simplify notations, we assume the electrolyte solution
contains a single species of cations and anions, with charges $\pm e$, and that 
they share the same diffusion coefficient $D$. In the online SI, we prove that 
our results continue to hold for the cases of asymmetric ions, as long as the 
fluid is charge-neutral when $E_0=0$.

The electric potential throughout the system is governed by Poisson's equation.
The solids are uniform, so the potential inside them obeys Laplace's Equation.  
In the solution, Poisson's equation reads
\begin{equation}
\nabla ^2 \psi( \vec{r},t)= -\frac{\rho( \vec{r},t)}{\epsilon_w' \epsilon_0};
\,\,\,\,\,\,\,\,
\rho ( \vec{r},t)= e (N_+( \vec{r},t) - N_-( \vec{r},t))
\label{poisson}
\end{equation}
Here $N_{\pm}$ are the number densities of the ions in the solution, $\psi$ is 
the potential in the fluid and $\epsilon_w'$is the \emph{static} relative 
permittivity of water.

The motion of ions in the liquid is characterized by three density currents: a
diffusive current driven by ion density gradients, a conductive driven by the 
electric field and a hydrodynamic current from ions being carried by the 
macroscopic motion of the solution itself.

Without significant $\zeta$-potential and surface charge on solids, the net 
charge density in the solution is due entirely to the external field.  Thus, 
the electrokinetic flow given by the Helmholtz-Smoluchowski 
equation~\cite{Levich} is proportional to $E_0^2$.  In the online SI we show 
that the ionic current carried by such an electrokinetic flow is much smaller 
than the conductive current, which is proportional to $E_0$, and can be safely 
ignored in the low-frequency and low field linear limit.

Thus, the ionic currents in the solution consist of diffusive and conductive 
parts, related by the Einstein relation. The ionic current densities and the 
corresponding ion number conservation equations are:
\begin{equation}
	\vec{j}_\pm^{\,N} =-D\left(\overrightarrow{\nabla} N_\pm \pm\frac{e 
N_\pm}{k_BT} \overrightarrow{\nabla}\psi\right),\,\,\,\,
{\overrightarrow\nabla}\cdot\vec{j}_{\pm}^{\,N} ( \vec{r},t)
= -{\partial N_{\pm}( \vec{r},t) \over \partial t} 
\label{charge_consv}
\end{equation}

Combining the Poisson Equation Eq.~\ref{poisson} and charge conservation
Eq.~\ref{charge_consv} produces three coupled non-linear partial
differential equations for $N_\pm$ and $\psi_L$. At a small enough driving 
field $E_0$ we can divide the ion densities into a background charge density
$N^0_\pm=N$ for the ions at $E_0=0$ and small perturbations due to $E_0$: 
$N_\pm=N+n_\pm$. Two crucial assumptions, that $n_\pm\ll N$ and 
$\overrightarrow{\nabla}N\approx0$, the latter due to the smallness of the
$\zeta$-potential and the insignificance of the double layer when $E_0=0$, lead 
to the linearization and simplification of the Eq.~\ref{charge_consv}:
\begin{equation}
\label{eq:motion}
{\overrightarrow\nabla}\cdot\vec{j}_{\pm}=i\omega n_{\pm},
\vec{j}_\pm =-D\left(\overrightarrow{\nabla} n_\pm \pm\frac{e N_0}{k_BT} 
\overrightarrow{\nabla}\psi\right),
\nabla^2 \psi=-\frac{e(n_+-n_-)}{ \epsilon_0 \epsilon_w'}.
\end{equation}
We show in the online SI that the condition for linearization is simply $e E_0 
a\ll k_B T$, where $a$ is the linear dimension
of the solid particles. This condition means, roughly, that the ion densities 
are not changed greatly from equilibrium values by the driving field, similar 
in spirits to the condition listed above for ignoring the $\zeta$-potential.  
Even for mm-sized particles, this condition is easily satisfiable.

For the boundary conditions on a solid-solution boundary $\mathbf{\Sigma}$, 
either the ideally polarized or the non-polarizable Butler-Volmer~\cite{Bard}/ 
Chang-Jaffe~\cite{Jaffe} boundary conditions (BCs) can be adopted. In this 
paper we study the former, and in another work we show that results here can be 
generalized to the latter.  We thus assume ions neither penetrate nor undergo 
redox reactions on the interface (ideally polarizable), and thus the normal 
ionic current vanishes on $\mathbf{\Sigma}$:
\begin{equation}\label{eq:bc}
\widehat{u}\cdot j_\pm |_{\mathbf{\Sigma}}
=0\,\mathbf{[a]},\,
\psi_{S}=\psi\,\mathbf{[b]},\,\,
\widehat{u}\cdot\overrightarrow{\nabla}\psi_S=
c\,\widehat{u}\cdot\overrightarrow{\nabla}\psi\,\mathbf{[c]},\,
c=\frac{i\omega  \epsilon _w'\epsilon _0}
{\sigma _{s}+i \omega \epsilon_0\epsilon_{s}'}
\end{equation}
Here we also list the standard BCs matching the potentials in the solid 
$\psi_S$ and in the solution $\psi$. Here $\widehat{u}$ is the
normal unit vector on the interface $\mathbf{\Sigma}$, and
$\sigma_s$ and $\epsilon_s'$ are the \emph{static} conductivity and relative 
permittivity of
the solid. We assume that in the low-frequency regime considered in this paper, 
the real dielectric constants and the conductivities of both the solid and the 
solution are frequency-independent.

Although, as mentioned above, the symmetry between anions and cations 
(\emph{e.g.} having the same mobilities) is not essential to our results, it 
does afford a further simplification to the governing equations 
Eq.~\ref{eq:motion} and the BCs, as the dynamics of the total ion density 
$n^{\mathrm{total}}=n_+ +n_-$ decouple from the driving field and it remains 
zero throughout the liquid and all the time.  The net ion density  follows a 
particularly simple equation, with a simple Green's function:
\begin{equation}\label{eq:net_motion}
	\nabla^2 n^{\mathrm{net}}=\beta^2 
	n^{\mathrm{net}};\,\,n^{\mathrm{net}}=n_+-n_-;\,\,
	G(\vec{r},\vec{r}\,')=-\frac{1}{4\pi}\frac{e^{-\beta|\vec{r}-\vec{r}\,'|}}{|\vec{r}-\vec{r}\,'|}.
\end{equation}
Here we introduce the Debye length $\lambda$ that sets the scale for 
electrostatic screening in the solution, and a characteristic time scale
$\tau_D$:
\begin{equation}\label{debye}
	\beta^2 \lambda ^2 = 1 +i \,\omega \tau_D\,;\,\,\,\,
\lambda^2 = \frac{k_BT\,\epsilon_0 \epsilon_w'}{2 N e^2};\,\,\,
\tau_D=  \frac{ \lambda^2}{D}=\frac{\epsilon_0\epsilon_w'}{\sigma_w}.
\end{equation}
Typical values of $\lambda$ range from 0.3\,nm to 10\,nm from concentrated 
($1M$) to dilute ($10^{-3}M$) electrolyte solutions, and $\tau_D$ ranges from 
$10^{-10}$s to $10^{-7}$s.  In this paper we consider low-frequency 
$\omega\tau_D\ll1$, so   $\beta\approx1/\lambda$.

\section{A Dilute Suspension of Spheres: Exact Solution}
For a single spherical particle of radius ``$a$'', an exact solution can be 
found. Polarization $P$, defined by the far field potential $\psi(|\vec{r}|\gg 
a)\to P\,a^3\,\vec{E_0}\cdot\vec{r}/(|\vec{r}|^3)-\vec{E}_0\cdot\vec{r}$, has a 
distinctive low frequency limit:
\begin{equation}
\label{eq:sphere-cond}
	\omega\tau_D\ll\frac{\lambda}{a}\frac{\sigma_{s}}{\sigma_w}, 
	\,\,\,\,\,\omega\tau_D\ll\frac{\lambda}{a},\,\,\,\,
	P\approx-\frac{1}{2}+\frac{3}{4}\frac{a}{\lambda}(i\omega\tau_D)
\end{equation}
According to  Maxwell-Garnett~\cite{Maxwell,MaxwellGarnett,Landauer,Batchelor} 
effective medium approximation, for a dilute suspension of volume fraction 
$f\ll1$, such that $f\,|P|\ll1$, the effective complex dielectric constant 
$\epsilon_{\mathrm{eff}}$ obeys the Clausius-Mossotti 
relation~\cite{Landauer,Batchelor}:
{\setlength\arraycolsep{0.5pt}
\begin{eqnarray}\label{Clausius}
f\,P&=&\frac{\epsilon_{\mathrm{eff}}/\epsilon_w(\omega)-1}{\epsilon_{\mathrm{eff}}/\epsilon_w(\omega)+2},\,\,\,
\epsilon_w(\omega)=\epsilon_w'+\frac{\sigma_w}{i\omega\epsilon_0},\,\,\,
\frac{\epsilon_{\mathrm{eff}}}{\epsilon_w(\omega)}\approx1+3f\,P.\\
\epsilon^{\prime}_{\mathrm{eff}}&\approx&
\epsilon_w'(1+3 f P^{\prime})+3f\frac{\sigma_w}{\omega\epsilon_0} P'',\,\,
\tan\theta(\omega)\stackrel{\mathrm{def}}{=}
	\frac{\epsilon'_{\textrm{eff}}}{\epsilon''_{\textrm{eff}}}
	\approx\frac{\epsilon'_{\textrm{eff}}}{\sigma_w/(\epsilon_0\omega)}.
\label{effective-def}
\end{eqnarray}}$P',P''$ are the real and imaginary parts of $P$.  
$\epsilon_{\mathrm{eff}}',\epsilon_{\mathrm{eff}}''$ are the real and imaginary 
parts of $\epsilon_{\mathrm{eff}}$. We also define here the phase shift 
$\theta(\omega)$ and use the fact that $\epsilon_{\mathrm{eff}}''$ is dominated 
by the solution's $\epsilon_w''=\sigma_w/\omega\epsilon_0$.

Now, using the fact for electrolyte solution,
$\sigma_w=\epsilon_0\epsilon_w'/\tau_D$, the real effective dielectric constant 
and the phase shift of the mixture is:
\begin{equation}
\label{eq:effectivefinal}
\frac{\epsilon^{\prime}_ {\mathrm{eff}}(\omega\to0)}{\epsilon_w'}
=1-\frac{3}{2}f+\frac{9}{4}\frac{a}{\lambda} f,\,\,
	\tan{\theta}=
	\frac{9f(\lambda/a)\omega\tau_D}
{(2\lambda/a)^2+(\omega\tau_D)^2(2\sigma_w/\sigma_s+1)^2}
\end{equation}
We see here that the imaginary part of the induced polarization $P$ generates 
an enhancement $a/\lambda$ to the dielectric constant of the mixture. For 
spheres whose $a\gg\lambda$, this enhancement, dependent only on their size, is 
very large.  However, to observe such enhancement for larger spheres, one has 
to go to lower frequencies controlled by factor $\lambda/a$.  Finally, if the 
particles are insulating, $\sigma_s=0$, enhancement cannot be observed, as the 
first inequality in Eq.~\ref{eq:sphere-cond} can never be satisfied.

The phase shift expression in Eq.\,\ref{eq:effectivefinal}, also valid beyond 
the low-frequency of Eqs.~\ref{eq:sphere-cond}, has a \emph{size-independent}
maximum of $(9/4)f/(1+2\sigma_w/\sigma_s)$ at 
$\omega\tau_D=(2\lambda/a)/(1+2\sigma_w/\sigma_s)$. For $\sigma_s\gg\sigma_w$, 
the maximum of $9f/4$ is easily observable by a commercial instrument, often 
with $0.1$-milirad sensitivity, even for a very small volume fraction $f$.
\section{Scaling Law of Universal Dielectric Enhancement}
We now generalize the results for the dielectric enhancement of spheres in 
electrolyte solution,
Eqs.\,\ref{eq:sphere-cond},\,\ref{eq:effectivefinal}, to particles of arbitrary 
shapes, which we state in the form of three propositions that establish
the universality, the scaling law and conditions for the enhancement. A final 
proposition demonstrates a scale-invariant maximum in the phase shift, easily 
accessible in impedance measurements. 

\noindent \textit{\textbf{Proposition 1.}} \textit{For a homogeneous
particle of any smooth, convex shape immersed in an electrolyte solution, 
driven by an external field
$E_0 \widehat{z} e^{i\omega t}$, as $\omega\to0$ the total dipole moment, 
measured far from
the particle, has the form 
$(\overline{\overline{P}}_r+(a/\lambda)i\omega\tau_D~\overline{\overline{P}}_i 
)~\vec{E}_0~V$.  Here
$\overline{\overline{P}}_r$ and $\overline{\overline{P}}_i$ are dimensionless 
rank two tensors depending solely on the shape,
but not on the size or the electrical properties of the particle.  ``$a$''
is the linear dimension of the particle and $V$ is its volume.}

Proposition I generalizes the low frequency limit Eq.~\ref{eq:sphere-cond} of 
spherical particles. A natural corollary of it, which generalizes 
Eq.\,\ref{eq:effectivefinal}, gives the dielectric enhancement for a dilute,
randomly oriented suspension of such particles under the effective medium 
theory.

\noindent\textit{\textbf{Corollary I.}} \textit{A dilute random
suspension of identical particles in an electrolyte solution
with volume fraction $f$
will have a geometrically enhanced low-frequency real dielectric
constant independent of the electrical properties of the particles:
$\epsilon_{\mathrm{eff}}'=\epsilon_w'(1+f(P_{r}+(a/\lambda)P_i))$.}

$P_r$ and $P_i$ are dimensionless numbers depending solely on the particles' 
shape, related to $\overline{\overline{P}}_r$ and $\overline{\overline{P}}_i$ 
in Proposition I by geometry.

The derivation of Corollary I from Proposition I is entirely analogous to the 
derivation Eq.~\ref{Clausius} from Maxwell-Garnett 
theory~\cite{Landauer,Batchelor}. The only significant difference is that the 
particle's shape being arbitrary, the polarization is no longer always colinear 
with the external field. This complication only introduces extra factors 
depending only on a particle's shape~\cite{Landauer}  but does not change the 
scaling behavior of the imaginary part of the dipole moment, identical to that 
in Eq.~\ref{eq:sphere-cond} which leads to dielectric enhancement in 
Eq.~\ref{eq:effectivefinal}. Random orientation of the particles ensures 
$\epsilon'_{\mathrm{eff}}$ is a scalar. 

The next proposition sets the conditions for observing the dielectric 
enhancement in Proposition I and Corollary I. They are generalizations of two 
inequalities in Eq.~\ref{eq:sphere-cond} of spherical particles.

\noindent\textbf{\textit{Proposition II.}} \textit{The conditions for observing 
the low-frequency dielectric enhancement in Proposition I and Corollary I 
are:\\
\begin{subequations}
\begin{tabularx}{\linewidth}{@{}XX@{}}
\begin{equation}\label{enhancement_condition_material}
	\omega\tau_D\ll\frac{1}{R}\frac{\lambda}{a}\frac{\sigma_{s}}{\sigma_w},
\end{equation}
&
\begin{equation}\label{enhancement_condition_geometry}
	\omega\tau_D\ll\frac{1}{R}\frac{\lambda}{a}.
\end{equation}
\end{tabularx}
\end{subequations}\\
$R$ is a dimensionless number, often of order one, determined solely by the 
shape of the particle. Its precise form will be shown in Appendix.}

The scale-invariant maximum in the phase shift of Eq.~\ref{eq:effectivefinal} 
can also be generalized to particles of arbitrary shapes:

\noindent\textbf{\textit{Proposition III.}} \textit{When $\sigma_w\ll\sigma_s$,
the phase shift of a mixture in Corollary I, $\theta(\omega)$ as defined in 
Eq.~\ref{effective-def}, has a maximum 
$\tan\theta_{\mathrm{max}}=f\,\mathsf{\Theta}$. $\mathsf{\Theta}$ is a 
dimensionless number determined solely by the particle's shape.}

Next, we develop a physical picture of an Externally Induced Double Layer and 
translate it into two effective boundary conditions.  We then use them to prove 
above propositions for arbitrary shapes.
\section{Separation of Scales and Effective Boundary Conditions}
In this section, we use two separations of scales, that the particle is large 
$a\gg\lambda$ and that the frequency is low $\omega\tau_D\ll1$, to encode all 
the ion charge dynamics into two effective boundary conditions relating the 
solid phase and the charge neutral liquid phase.

Given that for $\omega\tau_D\ll1$, 
$\beta=(1+i\omega\tau_D)^{1/2}/\lambda\approx1/\lambda$, the Green's function 
in Eq.~\ref{eq:net_motion} shows that any electrolyte
net charge disturbance due to an introduced charge decays within a nano-scale
length $\lambda$. Now consider any geometry where the minimal radius of 
curvature $R_0$
of any interface between the solid and the electrolyte solution is
still much larger than $\lambda$.  Under the driving field $E_0$, the induced
surface charges on the immersed bodies will in turn induce in the solution
a \emph{thin carpet of net charge} hugging the interface on the solution side,
with thickness on the order of a few $\lambda$.  We term this thin charged 
layer the ``\underline{E}xternally \underline{I}nduced \underline{D}ouble 
\underline{L}ayer'' (EIDL), to distinguish it from the intrinsic double layer 
in the presence of a $\zeta$-potential. Bazant and his
co-workers used a similar term in another context~\cite{Bazant}. The rest of 
the solution outside the EIDL will stay \emph{charge neutral}. 

For the rest of this paper, we use the subscript ``$L$'' for variables in the 
charge neutral-liquid \emph{outside} the EIDL, and subscript ``$S$'' for 
variables \emph{inside} the solids. Variables within the EIDL have no 
subscripts.

Given the large radius of curvature $R_0\gg\lambda$, all spatial variations
within the EIDL is much faster in the normal direction
$\widehat{u}$, so in the tangential directions the geometry can be modeled as a 
flat infinite plane. Other authors used similar approximations in finite 
$\zeta$-potential models~\cite{Chew,Fixman,Hinch}. By Eq.~\ref{eq:net_motion}, 
the net charge density in EIDL is
\begin{equation}\label{carpet_net}
	n^{\mathrm{net}}(\xi)=n_0 e^{-\beta\xi},
\end{equation}
where $n_0$ is the net charge density right at the particle-solution interface 
and we set $\xi$-axis in the normal direction $\widehat{u}$ into the solution.

The electric field within the EIDL will vary much more rapidly in the normal 
than in the tangential direction, with the former typically on the scale 
$1/\lambda$ and latter $1/R_0$. Thus, the Poisson equation in the layer 
involves only the normal component of $\vec{E}$:
\begin{equation}\label{carpet_poisson}
	\overrightarrow{\nabla}\cdot\vec{E}\approx\frac{dE^{\bot}}{d\xi}=\frac{en^{\mathrm{net}}(\xi)}
	{\epsilon_0\epsilon_w'},\,\,\,
	E^{\bot}(\xi)=E^{\bot}_{L}-\frac{e}{\epsilon_0\epsilon_w'}\int_{\xi}^{\infty} 
n^{\mathrm{net}}(\xi') d\xi'.
\end{equation}
Here $E^{\bot}_{L}$ is the normal field in the charge-neutral fluid \emph{just 
outside }the EIDL, corresponding to $\xi=+\infty$ since it lies at a distance 
$\gg\lambda$ away from the solid-solution interface.

According to Eq.~\ref{charge_consv}, the net ion current density, predominantly 
in the normal $\widehat{u}$ direction within the EIDL, can be written as
\begin{equation}
	j^{\,\mathrm{EIDL}}=-D\frac{dn^{\mathrm{net}}}{d\xi}+\frac{\sigma_w}{e} 
	E^{\bot}
	=\underbrace{-D\frac{dn^{\mathrm{net}}}{d\xi}
	-\frac{\sigma_w}{\epsilon_0\epsilon_w'}\int_{\xi}^{\infty} 
	n^{\mathrm{net}}(\xi') d\xi'}_{\textstyle j^{\,\mathrm{var}}}
	+\underbrace{\frac{\sigma_w}{e} E^{\bot}_{L}}_{\textstyle 
j^{\,\mathrm{out}}}.
\end{equation}
The current in the EIDL is broken into two parts. One is the rapidly varying 
part $j^{\,\mathrm{var}}$, due
to the net ion density gradient and the electric field generated by the net 
charges in the EIDL. The other is a spatially constant, divergence-free current 
$j^{\,\mathrm{out}}$ due solely to the $E^{\bot}_{L}$ carried over from the 
charge neutral, conductive liquid outside.

Since $n^{\mathrm{net}}(\xi)$ vary spatially as $e^{-\beta\xi}$, as does its 
spatial derivatives and integrals, so
does the rapidly varying current $j^{\,\mathrm{var}}$ due to them. Thus, charge 
conservation gives a starkly simple relation between the charge and current 
density:
\begin{equation}\label{net-charge-conserv}
	\overrightarrow{\nabla}\cdot j+{\partial n^{\mathrm{var}}\over \partial 
	t}
	\approx \frac{dj^{\,\mathrm{var}}}{d\xi}+{\partial 
	n^{\mathrm{net}}\over \partial t}=
	-\beta j^{\,\mathrm{var}}+i\omega n^{\mathrm{net}}=0 
	,\,\,j^{\,\mathrm{var}}=\frac{i\omega}{\beta} n^{\mathrm{net}}.
\end{equation}
Naturally the spatially uniform current from the neutral liquid 
$j^{\,\mathrm{out}}$ does not contribute to  the divergence of net particle 
current.

We can now relate the net charge distribution in the EIDL 
$n^{\mathrm{net}}(\xi)=n_0 e^{-\beta\xi}$ to the electric field outside in the 
charge neutral liquid, by applying the crucial blocking boundary condition 
Eq.~\ref{eq:bc}a that the net current vanishes at the solid-solution interface:
\begin{equation}\label{blocking_BC}
	j^{\,\mathrm{var}}(\xi=0)+j^{\,\mathrm{out}}=0,\,\,\,\,n_0=-\frac{E^{\bot}_{L}\beta\sigma_w}{i\omega 
e}.
\end{equation}

Now, integrating Poisson's equation in the EIDL Eq.~\ref{carpet_poisson} and 
using the charge distribution Eq.~\ref{carpet_net}, we can obtain the normal 
electric field at $\xi=0$, on the solution side:
\begin{equation}\label{carpet_surface_field}
	E^{\bot}(\xi=0)
	=E^{\bot}_{L}-\frac{e}{\epsilon_0\epsilon_w'}\int_{0}^{\infty} n(\xi') 
	d\xi'
	=\left(1+\frac{1}{i\omega\tau_D}\right) E^{\bot}_{L}.
\end{equation}
A further integration of Poisson's equation gives the potential drop across the 
EIDL:
\begin{equation}\label{carpet_potential_drop}
	\psi_{L}-\psi(\xi=0)=-\int_0^{d} d\xi'E^{\bot}(\xi)
	=-E^{\bot}_{L}\left(d +\frac{1}{i\omega\tau_D\beta}\right).
\end{equation}
Here $d$ is the ``thickness'' of the EIDL. Because the net charged 
density decays exponentially in a few $\lambda$, and the frequency 
$\omega\ll1/\tau_D$, we have 
$1/(\omega\tau_D\beta)\approx\lambda/(\omega\tau_D)\gg d$, so we can drop $d$.

Combining Eq.~\ref{carpet_surface_field} and Eq.~\ref{carpet_potential_drop}, 
and using the boundary condition between the fluid and interface 
$\psi_{S}=\psi(\xi=0)$ and $E^{\bot}_{S}=c E^{\bot}(\xi=0)$ 
(Eq.\,\ref{eq:bc}b,c), we obtain the following effective boundary conditions 
(BCs) relating
the fields and potentials \emph{inside the solids} \emph{directly} to those in
the \emph{charge neutral liquid outside EIDL}:
\begin{equation}\label{BC_eff}
	E^{\bot}_{S}=c\left(1+\frac{1}{i\omega\tau_D}\right)E^{\bot}_{L}
	\,\,\,\,\,\,\mathbf{(\!a)},\,\,\,\,\,
	\psi_{S}=\psi_{L}+\frac{1}{i\omega\tau_D\beta } E^{\bot}_{L}.
\,\,\,\,\mathbf{(\!b)},
\end{equation}
The BCs Eqs.\,\ref{BC_eff} greatly simplify the original equations of motion in 
the first section.  As the solid phase is electrically homogeneous, these BCs 
reduce coupled  Poisson and Helmholtz equations in Eq.~\ref{poisson} and 
Eq.~\ref{eq:net_motion} to two independent Laplace equations.  All the charge 
dynamics in the electrolyte solution, which is located exclusively in the EIDL, 
have been encoded in these two BCs. This enables us to apply results from the 
theory of harmonic functions to a solid-solution system, which we will resort 
to below.

The physics of Eq.~\ref{BC_eff}a is made transparent by rewriting it as 
\begin{equation}\label{equ:BC_eff_rewrite}
	\frac{E^{\bot}_{S}}{E^{\bot}_{L}}
	=c\left(1+\frac{1}{i\omega\tau_D}\right)
	=\frac{\epsilon_0\epsilon_w'+\sigma_w/(i\omega)}{\epsilon_0\epsilon_s'+\sigma_s/(i\omega)}
	=\frac{\epsilon_w(\omega)}{\epsilon_s(\omega)}.
\end{equation}
Here we use $\epsilon_{w,s}(\omega)$ to denote the frequency dependent complex
dielectric constant in the solution and in solids at low frequency. Thus the BC  
Eq.~\ref{BC_eff}a is nothing more than the generalization of the usual BC for 
normal AC electric fields on conductive boundaries, incorporating displacement 
currents, derived from charge conservation, with a form of conductivity proper 
to the electrolyte solution.

Eq.~\ref{BC_eff}b, on the other hand, is clearly due to a dipole layer whose 
density is $-(\epsilon_0\epsilon_w'E^{\bot}_{L})/( i\omega\tau_D\beta)$, which 
simple algebras show is exactly the effective dipole density of the EIDL
$\int e\,n^{\mathrm{net}}(\xi')\xi' d\xi'$. The difference between this dipole 
layer and the Gouy-Chapman double layer is simply that the dipole moment here 
is purely induced by, and is therefore proportional to, the external drive 
$E_0$.
\section{Size and Frequency Scaling: Dimensional Arguments}
In this section, we give qualitative arguments, based purely on the analytical 
properties of dielectric responses and the form of the effective boundary 
conditions Eqs.~\ref{BC_eff}a,\ref{BC_eff}b, that severely constrain the form 
of the low-frequency limit of the dipole moment. The same arguments also show 
the scale invariance of the impedance phase shift maximum. We then outline a 
perturbation theory that computes this dipole moment but leave the mathematical 
details to the Appendix.

As is known from elementary electrodynamics, the fact that dielectric responses 
must be real in the time domain requires that the real part of the complex 
dielectric response function to be symmetric and the imaginary part
antisymmetric under $\omega\to-\omega$. Furthermore, as the hydrodynamic flow 
is small and we have linearized the governing equations Eq.~\ref{eq:motion}, 
the induced dipole moment will be purely linear to the external
drive $E_0$. Thus the most general form of the total dipole moment as 
$\omega\to0$ is $(\widetilde{P}_r+ i\omega\widetilde{P}_i)E_0$, with 
$\widetilde{P}_i,\widetilde{P}_j$ independent of $\omega$.

Furthermore, when
$\omega\tau_D\ll(\epsilon_w'/\epsilon_{s}')(\sigma_{s}/\sigma_w)$,
the boundary condition Eq.~\ref{eq:bc}c
becomes dominated by conduction currents, and $c\approx i\omega\tau_D
(\sigma_w/\sigma_{in})$ becomes purely imaginary. As $\omega\tau_D\ll1$, the 
effective boundary conditions Eq.~\ref{BC_eff} take
a starkly simple form:
\begin{equation}\label{BC_eff_low_freq}
	E^{\bot}_{S}=\frac{\sigma_w}{\sigma_{s}}E^{\bot}_{L}
	\,\,\,\,\,\,\mathbf{(\!a)},\,\,\,\,\,
	\psi_{S}=\psi_{L}+\frac{1}{i\omega\tau_D\beta}E^{\bot}_L\,\,\,\,\,\,\mathbf{(\!b)},
\end{equation}
In this limit, in its \emph{only} occurrence in the governing Laplace equation 
and BCs, $\omega$ appears
in the combination $\omega\tau_D\beta\approx\omega\tau_D/\lambda$.
Thus precisely
this combination will appear in the imaginary term linear to $\omega$ in the 
total polarization $P$. Assume that the maximum of $\tan\theta(\omega)$ 
\emph{exists}, which we will show in Appendix below, it will clearly have no 
dependence on $\omega$.  Yet the \emph{only} occurrence of $\lambda$ appears 
precisely in the \emph{only} combination where $\omega$ occurs: 
$\omega\tau_D/\lambda$.  Thus $\tan\theta_{\mathrm{max}}$,
cannot have dependence on $\lambda$, either.

Finally, if we enlarge the linear dimension of the particle by a factor of 
$\alpha$ while preserving its shape, the boundary conditions, and thus the 
fields will remain the same provided we also rescale $\lambda$ by $\alpha$. 
Thus, the total dipole moment scales as the volume of the particle, and there 
is a prefactor $a$ before $i\omega\tau_D/\lambda$ that scales linearly with the 
size of the particle, rendering the combination dimensionless. So the most 
general form of the total dipole moment as $\omega\to0$ is
$(P_r+(a/\lambda)i\omega\tau_D\,P_i)VE$, by virtues of the conduction dominated 
form of Eq.~\ref{BC_eff_low_freq}, and little more than dimensional analysis 
beyond that.

$\mathrm{tan}\,\theta_{\mathrm{max}}$, determined solely by the polarization 
$P$ (Eq.\,\ref{effective-def}) but having no dependence on $\lambda$, will 
depend on size ``$a$'' only through volume $V$, which is absorbed into $f$. So 
$\mathsf{\Theta}$ in Proposition III is scale-invariant.

What remains to be shown is that the prefactors $\mathsf{\Theta},P_i,P_r$
does not depend on the material parameter $\sigma_w/\sigma_s$.  We also need to 
answer the physically crucial question of the \emph{conditions} under which the 
enhancement in Proposition and Corollary I are realized. One such condition 
Eq.\,\ref{enhancement_condition_material}, for example, shows that insulating 
particles with $\sigma_s=0$ cannot produce enhancement in our model without 
$\zeta$-potential. 

The EIDL dipole moment in Eq.~\ref{BC_eff}b, proportional to $E^{\bot}_L$, is 
purely imaginary and out of phase. It becomes dominant at low frequency and is 
the cause of the dielectric enhancement there. Introduce a length scale 
$aR_S=|\psi_S/\widehat{u}\cdot\overrightarrow{\nabla}\psi_{S}|$.  Assume the 
maximum of $R_S$ across the surface is around $R$. When $|\omega\tau_D\beta a 
R|\ll 1$, the second term in Eq.~\ref{BC_eff_low_freq}b dominates, making the 
EIDL dipole moment the main contributor to dielectric responses. This is partly 
the physics behind the condition Eq.~\ref{enhancement_condition_geometry} of 
Proposition II.  Furthermore, if particles are insulating, $\sigma_s=0$ forces 
the normal field $E_{L}^{\bot}$ to zero, \emph{eliminating} EIDL dipole moments 
altogether, partly explaining the condition 
Eq.\ref{enhancement_condition_geometry}.

To make comparisons of scales like these more precise and useful, we
use the Philips integral formulation of electrostatics~\cite{Philips} to 
combine Laplace equations and effective boundary conditions 
Eq.~\ref{BC_eff_low_freq} into two \emph{self-contained} integral equations 
\emph{on the interface} between the solid and the charge neutral liquids. We 
then find a perturbative solution for the $\psi_L$ integral equation at the 
$\omega\to0$ limit and use Green's theorem to compute the dipole moment 
explicitly from $\psi_L$ on the interface.  The conditions for the perturbative 
expansion are precisely those in Proposition II and the dipole moment has the 
form in Proposition I. The details can be found in the Appendix below, where we 
also show the \emph{existence} of a maximum in the phase shift 
$\tan\theta(\omega)$.
\section{Discussions}
In this work we show that dielectric enhancement can happen in a mixture of 
non-insulating particles and electrolyte solution, with no $\zeta$-potential on 
their ideally polarized, non-reactive interfaces. We also derive a universal 
scaling law for the magnitude of such enhancement. Our results comport well 
with preliminary data from recent experiments, which show clear size-dependent 
dielectric enhancements in a conductive material with non-reactive surfaces.  
Furthermore, we show that if particles are much more conductive than the 
solution, for generic shapes there is always a \emph{scale-invariant} maximum 
in the frequency-dependent phase shift in impedance measurements.  Evidence of 
this is also observed in the preliminary data.

Beyond direct relevance to these experiments, our theory also opens up a wider 
range of candidates for applications involving dielectric enhancements.  For 
example, materials of smaller $\zeta$-potential can now be considered for 
inclusions in electrolyte solutions as contrast agents, particularly where 
electrochemically inert interfaces are advantageous, such as in biological 
applications.  Furthermore, the scale invariance of the phase shift maximum 
means even sub-micron particles can produce large phase shifts at low volume 
fraction, opening the application of dielectric enhancement to intracellular 
probes.

The geometric scaling for the magnitude of dielectric enhancement may also 
prove useful for characterizing the statistical properties of the particles in 
complex fluids. For example, as we will show in another publication, the 
frequency dependence of the phase shift in a system of inhomogeneous spherical 
particles in solution can provide revealing information about the size 
distribution the particles.

Finally, the methods we develop here to prove the universal scaling of 
dielectric enhancement for arbitrary particle shapes may prove useful for 
studying other effects of interfacial processes in an electrolyte solution 
beyond the problem of dielectric enhancement. With a physical picture of 
externally induced dielectric enhancement (EIDL), we derive a pair of boundary 
conditions that are applicable to any interface that has small 
$\zeta-$potential, not just that of particular suspensions. Examples of such 
interfaces include the oil in water emulsions in hydrocarbon productions or 
porous rocks soaked in saline solution.  These boundary conditions encode all 
the complex charge dynamics on the interfaces, and reduce the physics away from 
the interfaces to that of simple charge-free electrostatics, tractable by a 
rich variety of traditional analytical and computational methods.


\section{Appendix: Integral Equation Perturbation Theory}
Here we tighten the scaling arguments above and derive a perturbative solution 
that shows prefactors $\overline{\overline{P}}_r,\overline{\overline{P}}_i$ in 
Proposition I to be material-independent.  The two inequalities in 
Eqs.\,\ref{enhancement_condition_material},\ref{enhancement_condition_geometry} 
of Proposition II come about naturally as \emph{conditions} for the 
perturbative expansions.

As boundary conditions Eqs.\,\ref{BC_eff} incorporate all the free charge 
dynamics in the
system, the potentials in the solid and the neutral liquid, on
either side of the EIDL, obey Laplace equation and are \emph{harmonic 
functions}.  Using Stokes' Theorem,
we can incorporate the BCs Eqs.\,\ref{BC_eff} to compute the potential at 
\emph{any point $\vec{r}$ in space} solely from its \emph{surface 
integrals}~\cite{Philips,Bladel}, see also online SI.
{\setlength\arraycolsep{0.5pt}
\begin{eqnarray}\label{eqn:potential-in-bulk}
	\widetilde{\epsilon}_{\mathsf{\Omega}}\psi_{\mathsf{\Omega}}(\vec{r})&=&\widetilde{\epsilon}_L\psi_0(\vec{r})\\
	&\mp&\frac{1}{4\pi}\oiint\limits_{\mathbf{\Sigma}}
	\left[\widetilde{\epsilon}_{\mathsf{\Omega}}\psi_{\mathsf{\Omega}}(\vec{r}\,'_{\mathbf{\Sigma}})-\psi^{\mathrm{eff}}_{\mathsf{\Omega}}(\vec{r}\,'_{\mathbf{\Sigma}})\right]\,
	\widehat{u}\cdot\overrightarrow{\nabla}_{\vec{r}\,'_{\mathbf{\Sigma}}}\frac{1}{|\vec{r}-\vec{r}\,'_{\mathbf{\Sigma}}|}dS',\nonumber
	\\
	\psi^{\mathrm{eff}}_{\mathsf{\Omega}}&=&\widetilde{\epsilon}_{-\mathsf{\Omega}}\psi_{\mathsf{\Omega}}\pm\frac{1}{i\omega\tau_D\beta}
	\widehat{u}\cdot\overrightarrow{\nabla}\psi_{\mathsf{\Omega}}.\label{effective_potential}
\end{eqnarray}}
Here we use subscript $\mathsf{\Omega}=S/L$ to label the Solid/(neutral) Liquid 
domain on either sides of the EIDL, which is considered infinitely thin and 
serves as boundary $\mathbf{\Sigma}$. $-\mathsf{\Omega}$ is the domain on the 
opposite side.  We use \emph{rescaled} dielectric constants 
$\widetilde{\epsilon}_S=1$ and 
$\widetilde{\epsilon}_{L}=c(1+1/(i\omega\tau_D))$, a result of this integral 
formulation~\cite{Bladel}. Under the mild assumption 
$\omega\tau_D\ll(\epsilon_w'/\epsilon_{s}')(\sigma_{s}/\sigma_w)$, 
$\widetilde{\epsilon}_{L}\approx\sigma_{w}/\sigma_{s}$ is real and frequency 
independent. $\psi_0=-E_0 z$ is the ``incident potential'' due to the driving 
electric field. The $\mp$ in Eq.~\ref{eqn:potential-in-bulk} and $\pm$ in 
Eq.~\ref{effective_potential}  are for $\psi_{S}$ and $\psi_{L}$ respectively.



Now, letting $\vec{r}$ approach $\mathbf{\Sigma}$. The integral in 
Eq.~\ref{eqn:potential-in-bulk} picks up new terms (the two terms before the 
$\pm$ below) from the discontinuity of 
$\overrightarrow{\nabla}(1/|\vec{r}-\vec{r}\,'|)$ and we 
have~\cite{Philips,Bladel}:
{\setlength\arraycolsep{0.5pt}
\begin{eqnarray}
	\widetilde{\epsilon}_L\psi_0(\vec{r}_{\mathbf{\Sigma}})&=&\frac{1}{2}\widetilde{\epsilon}_{\mathsf{\Omega}}\psi_{\mathsf{\Omega}}(\vec{r}_{\mathbf{\Sigma}})+
	\frac{1}{2}\psi^{\mathrm{eff}}_{\mathsf{\Omega}}(\vec{r}_{\mathbf{\Sigma}})\label{integral_eq}\\
	&\pm&\frac{1}{4\pi}\oiint\limits_{\mathbf{\Sigma}}
	\left[\widetilde{\epsilon}_{\mathsf{\Omega}}\psi_{\mathsf{\Omega}}(\vec{r}\,'_{\mathbf{\Sigma}})-\psi^{\mathrm{eff}}_{\mathsf{\Omega}}(\vec{r}\,'_{\mathbf{\Sigma}})\right]\,
	\widehat{u}\cdot\overrightarrow{\nabla}_{\vec{r}\,'_{\mathbf{\Sigma}}}\frac{1}{|\vec{r}_{\mathbf{\Sigma}}-\vec{r}\,'_{\mathbf{\Sigma}}|}dS',\nonumber
\end{eqnarray}}
Unlike Eqs.~\ref{eqn:potential-in-bulk}, Eqs.~\ref{integral_eq} are 
\emph{self-contained integral equations} that fully determine $\psi$ \emph{on 
the surface} $\mathbf{\Sigma}$, since both $\vec{r}_{\mathbf{\Sigma}}$ and 
$\vec{r}\,'_{\mathbf{\Sigma}}$ lie on $\mathbf{\Sigma}$. In discussions below, 
if all positions $\vec{r},\vec{r}\,'$ lie on $\mathbf{\Sigma}$ we omit 
subscript ${\mathbf{\Sigma}}$.  It is important to note that the separate 
integral equations for $\psi_{S}$ and $\psi_{L}$ do not couple to one another.  
All physical influences from  the other side of the EIDL have been encoded in 
the very form of $\psi_{\mathrm{eff}}$.

Rescaling the potential inside the solid $\psi_S'=\psi_S/(i\omega\tau_D\beta 
a\widetilde{\epsilon}_L)$, the integral equation Eq.~\ref{integral_eq} for 
$\psi_S$ can be rewritten as:
{\setlength\arraycolsep{0.0pt}
\begin{eqnarray}\label{eq:solid-int-rescale}
	&&\lefteqn{\psi_0(\vec{r})=
	\frac{1}{2}\left[(1+\widetilde{\epsilon}_L)(i\omega\tau_D a\beta)+
	\frac{1}{R_S(\vec{r})}\right]\psi_S'(\vec{r})}\\
	&&+\frac{1}{4\pi}\oiint\limits_{\mathbf{\Sigma}}
	\left[(1-\widetilde{\epsilon}_L)(i\omega\tau_D a\beta)
	-\frac{1}{R_S(\vec{r}\,')}\right]\psi_S'(\vec{r}\,')
	\widehat{u}\cdot\overrightarrow{\nabla}_{\vec{r}\,'}\frac{1}{|\vec{r}-\vec{r}\,'|}dS'.\nonumber
\end{eqnarray}
}
Here, we introduce a new dimensionless shorthand variable:
\begin{equation}\label{eq:R-def}
	R_S(\vec{r})=
	\frac{1}{a}\frac{\psi_S(\vec{r})}
	{\widehat{u}\cdot\overrightarrow{\nabla}\psi_S(\vec{r})}
	=\frac{1}{a}\frac{\psi'_S(\vec{r})}
	{\widehat{u}\cdot\overrightarrow{\nabla}\psi'_S(\vec{r})},\,\,\,
R=\max_{\vec{r}\mathbf{\in\Sigma}}(|R_S(\vec{r})|)
\end{equation}
For a spherical particle, $R_S=1$, but it generally varies across the 
interface.
For the simplicity of presentation, we first assume the normal electric field 
never vanishes across the surface $\mathbf{\Sigma}$:
$|E^{\bot}_S|=|\widehat{u}\cdot\overrightarrow{\nabla}\psi_S|>0$. Since the 
electric field inside the solid and the linear size of the particle are both 
finite, $\psi_S$ and $\psi_S'$ are everywhere finite on surface 
$\mathbf{\Sigma}$, so is $R_S$ defined by Eq.~\ref{eq:R-def}. (This does not 
hold if the solid domain extends infinitely, as in macroscopically extended 
porous media.)

Now set the $R$ in 
Eqs.\,\ref{enhancement_condition_material},\ref{enhancement_condition_geometry} 
as the maximum of $R_S$ across $\mathbf{\Sigma}$ (Eq.~\ref{eq:R-def} above),
simple algebra shows that if \emph{both} conditions Proposition II are 
satisfied, the $1/R_S$ term in Eq.~\ref{eq:solid-int-rescale} dominates the 
integrals, and $R_S$ can be self-consistently determined by a simpler equation:
{\setlength\arraycolsep{-0.0pt}
\begin{eqnarray}\label{equ:R-self-consistent}
	\psi_0(\vec{r})&=&
	\frac{1}{2}
	\frac{1}{R_S(\vec{r})}\psi_S'(\vec{r})-\frac{1}{4\pi}
	\oiint\limits_{\mathbf{\Sigma}}
	\frac{1}{R_S(\vec{r}\,')}\psi_S'(\vec{r}\,')
	\widehat{u}\cdot\overrightarrow{\nabla}_{\vec{r}\,'}\frac{1}{|\vec{r}-\vec{r}\,'|}dS'\nonumber\\
	&=&\frac{a}{2}\widehat{u}\cdot\overrightarrow{\nabla}\psi'_S(\vec{r})
	-\frac{a}{4\pi}\oiint\limits_{\mathbf{\Sigma}}
\widehat{u}\cdot\overrightarrow{\nabla}\psi'_S(\vec{r\,'})
	\widehat{u}\cdot\overrightarrow{\nabla}_{\vec{r}'}\frac{1}{|\vec{r}-\vec{r}\,'|}dS'
\end{eqnarray}
}Eq.~\ref{equ:R-self-consistent} clearly does not depend on the electrical 
properties of the solid particle or the liquid. Nor does it depend on the 
frequency $\omega$.  The right hand side is furthermore invariant under a 
rescaling of the linear dimension of the particle $a\to\alpha a$, so $\psi'_S$ 
is simply proportional to $\psi_0\propto E_0a$, with proportionality determined 
solely by particle shape.
Thus $R_S$, and by extension $R$ defined above, are determined solely by the 
\emph{shape} of the solid particle.

Furthermore, as Eq.~\ref{equ:R-self-consistent} is invariant with respect to 
rescaling $a$, and there is \emph{no extra parameters} in the integral 
equation
Eq.~\ref{equ:R-self-consistent} other than the particle's shape, we expect that 
$R_S$ will generally be of order one, as long as the particle's shape is smooth 
and convex, and its aspect ratios are near one (e.g. it is not shaped like a 
needle or a disc), so that there is no additional \emph{geometric} parameters 
much smaller or larger than one.

Even when there are points on the surface where $E^{\bot}_S$ vanishes, which is
the case for a generic particle shape, the arguments above can be modified to 
preserve the validity of the reduction of Eq.~\ref{eq:solid-int-rescale} to 
Eq.~\ref{equ:R-self-consistent}, as long as the places where $E^{\bot}_S=0$ is 
a measure zero subset of the surface $\mathbf{\Sigma}$, and near these points 
$E^{\bot}_S$ approaches zero rapidly.

To see this most simply, imagine we discretize the surface $\mathbf{\Sigma}$ 
and make $\psi_S'(\vec{r}\,')$ into a column vector.
The integral equation Eq.~\ref{eq:solid-int-rescale} will become a set of 
linear equations, with the kernel of the integral becoming a matrix and 
$\vec{r},\vec{r}\,'$ becoming column and row indices.

As $\omega\to0$, the only matrix elements that \emph{do not} trivially reduce 
to that of Eq.~\ref{equ:R-self-consistent} are the rows labeled
by $\vec{r}\,'$where $E^{\bot}_S$ approaches zero, so that $1/R_S(\vec{r}\,')$
becomes small. However, since the points where $E^{\bot}_S$ vanishes are a 
measure zero subset, and near them $E^{\bot}_s$ declines rapidly,
as $\omega\tau_D a\beta$ becomes smaller, the number of rows remaining 
significantly different between Eq.\,\ref{eq:solid-int-rescale} and 
Eq.\,\ref{equ:R-self-consistent} rapidly becomes a negligible portion of the 
total matrix.  So at a large enough $R$, which would be determined solely by 
the shape of the particle, the two integral kernels will converge, to the 
lowest order of parameter $\omega\tau_D a\beta$.

Having proved that $R_{S}$ at low-frequency are determined by particle shape 
alone, we now use the two conditions in Proposition II,
to define small parameters in a perturbative solution for the potential 
$\psi_L$, from which we derive an expression of the dipole moment in the form 
given in Proposition I and thus proving both propositions.

Combining the definition Eq.~\ref{eq:R-def} and effective BCs Eq.~\ref{BC_eff}, 
we can recast the effective potential $\psi^{\mathrm{eff}}_L$ in terms of 
$R_S$:
\begin{equation}\label{epsilon_eff_out}
	\psi^{\mathrm{eff}}_L
	=\left(1-\frac{1}{1+(i\omega\tau_D\beta a R_S)
	\widetilde{\epsilon}_L}\right)\psi_{L}
	\approx(i\omega\tau_D\beta a R_S\widetilde{\epsilon}_L)\psi_L.
\end{equation}
The last step uses an expansion of the fraction, which, as 
$\widetilde{\epsilon}_L\approx\sigma_w/\sigma_s$, is valid when 
Eq.~\ref{enhancement_condition_material} of Proposition II holds.  Of course, 
near the sets of point where $E^{\bot}_S=0$ and $R_S$ diverges, this 
approximation fail.  But as discussed above, when $\omega\tau_D\beta a$ becomes 
sufficiently small, such points constitute a insignificant portion of the 
integral kernel.

With $\,\,\omega\tau_D\beta a R\ll1$, which is the condition 
Eq.~\ref{enhancement_condition_geometry} in Proposition II, we can apply the 
following perturbative solution:
{\setlength\arraycolsep{0.0pt}
\begin{eqnarray}\label{eqn:perturbation}
	&&\lefteqn{\widetilde{\epsilon}_L\psi_0(\vec{r})=
	\frac{1}{2}\left(\widetilde{\epsilon}_L+
	i\omega\tau_D\beta a R_S\,\widetilde{\epsilon}_L\right)
	(\psi^0_L(\vec{r})+\delta\psi_L(\vec{r}))}\\
	&&-\frac{1}{4\pi}\oiint\limits_{\mathbf{\Sigma}}
	\left(\widetilde{\epsilon}_L-
	i\omega\tau_D\beta a R_S\,\widetilde{\epsilon}_L\right)
	(\psi^0_L(\vec{r}\,')+\delta\psi_L(\vec{r}\,'))
	\widehat{u}\cdot\overrightarrow{\nabla}_{\vec{r}\,'}\frac{1}{|\vec{r}-\vec{r}\,'|}dS',\nonumber
\end{eqnarray}}
The electrical property of solids $\widetilde{\epsilon}_L$ now cancels. The 
starting point $\psi^0_L$ is simply the solution of this zeroth order order 
integral equation:
\begin{equation}\label{eqn:zeroth-order}
	\psi_0(\vec{r})=
	\frac{1}{2}
	\psi^0_L(\vec{r})
	-\frac{1}{4\pi}\oiint\limits_{\mathbf{\Sigma}}
	\psi^0_L(\vec{r}\,')
	\widehat{u}\cdot\overrightarrow{\nabla}_{\vec{r}\,'}\frac{1}{|\vec{r}-\vec{r}\,'|}dS'
\end{equation}
$\psi^0_L$ obviously is purely real.  Since on the surface $\mathbf{\Sigma}$ 
the driving potential $\psi_0$ scales as $E_0a$, so does $\psi^0_L$.  
Furthermore, in Eq.~\ref{eqn:zeroth-order}, a rescaling of the particle size 
``$a$'' will only affect $\psi_0$, but not the integral kernel 
$\widehat{u}\cdot\overrightarrow{\nabla}_{\vec{r}\,'}\frac{1}{|\vec{r}-\vec{r}\,'|}dS'$,
so
$\psi^0_L(\vec{r})=E_0a\,p_r(\vec{r})$,
where $p_r(\vec{r})$ is a \emph{dimensionless} function determined 
\emph{solely} by the \emph{shape} of a particle.

We obtain the equation for the first order perturbation $\delta\psi_L$ by 
subtracting Eq.~\ref{eqn:zeroth-order} from Eq.~\ref{eqn:perturbation} and 
ignoring the second order terms:
{\setlength\arraycolsep{0.5pt}
\begin{eqnarray}\label{eqn:perturbation-solution}
	&&\delta\psi_L(\vec{r})
	-\frac{1}{4\pi}\oiint\limits_{\mathbf{\Sigma}}
	\delta\psi_L(\vec{r}\,')
	\widehat{u}\cdot\overrightarrow{\nabla}_{\vec{r}\,'}\frac{1}{|\vec{r}-\vec{r}\,'|}dS'\\
	&=&-i\omega\tau_D\beta a\left[
	R_S(\vec{r})\psi^0_L(\vec{r})
	+\frac{1}{4\pi}\oiint\limits_{\mathbf{\Sigma}}
	R_S(\vec{r}\,')\psi^0_L(\vec{r}\,')
	\widehat{u}\cdot\overrightarrow{\nabla}_{\vec{r}\,'}\frac{1}{|\vec{r}-\vec{r}\,'|}dS'
	\right].\nonumber
\end{eqnarray}}
Since $\psi^0_L$ is purely real, the imaginary part of the dielectric response 
$P_i$ necessarily comes from the purely imaginary $\delta\psi_L$. The part in 
the square bracket on the right hand side in 
Eq.~\ref{eqn:perturbation-solution} scales linearly with $\psi^0_L$ and thus 
with $E_0a$, but is otherwise entirely invariant to rescaling of particle 
dimension ``$a$''.  So, combined with the scaling for $p_r$ above:
\begin{equation}\label{eqn:P_scaling}
\psi_L(\vec{r})=
\psi^0_L(\vec{r})+\delta\psi_L(\vec{r})=E_0a\left[\,p_r(\vec{r})+\,(i\omega\tau_D\beta 
a)~p_i(\vec{r})\right],
\end{equation}
where $p_i(\vec{r}),p_r(\vec{r})$ are, again, \emph{dimensionless} functions 
determined \emph{solely} by the \emph{shape} of the particle.

Now we return to Eq.~\ref{eqn:potential-in-bulk}. When 
$|\vec{r}|\gg|\vec{r}\,'_{\mathbf{\Sigma}}|$,
$\overrightarrow{\nabla}_{\vec{r}\,'_{\mathbf{\Sigma}}}\frac{1}{|\vec{r}-\vec{r}\,'_{\mathbf{\Sigma}}|}\approx\frac{\vec{r}}{|\vec{r}|^3}$.
The expansion in Eq.~\ref{epsilon_eff_out} also greatly simplifies the surface 
integral:
\begin{eqnarray}\label{eqn:dipole}
	\psi_L(\vec{r})
	&=&\psi_0(\vec{r})+
	\frac{1}{4\pi\epsilon_0}\frac{\vec{r}\cdot\vec{P}}{|\vec{r}|^3}\\
	&=&\psi_0(\vec{r})
	+\frac{\vec{r}}{4\pi|\vec{r}|^3}\cdot\oiint\limits_{\mathbf{\Sigma}}
	\psi_{L}\left(1-i\omega\tau_D\beta a R_S\right)\,\widehat{u}\, 
	dS'\nonumber
	.
\end{eqnarray}
Thus, the dipole moment $\vec{P}$ can be simply read off, from the behavior of 
$\psi_L$ at large $\vec{r}$, as the surface integral in Eq.~\ref{eqn:dipole}.  
When the two conditions 
Eq.~\ref{enhancement_condition_material},\ref{enhancement_condition_geometry} 
in Proposition II are satisfied, $\psi_L\approx\psi^0_L+\delta\psi_L$. The real 
and imaginary part are in turns defined by integral equations 
Eq.~\ref{eqn:zeroth-order},\ref{eqn:perturbation-solution}, well-posed Fredholm 
equations of the second kind~\cite{Bladel}.

Applying the scaling behaviors of $\psi^0_L$ and $\delta\psi$ in 
Eq.\,\ref{eqn:P_scaling}, we have the following
scaling of the dipole integral in Eq.~\ref{eqn:dipole}:
\begin{eqnarray}\label{eqn:dipole-scaling}
	&&\oiint\limits_{\mathbf{\Sigma}}
	\psi_{L}\left(1-i\omega\tau_D\beta a R_S\right)\,\widehat{u}\, dS'\\
	&&\approx E_0 a\oiint\limits_{\mathbf{\Sigma}}\,
	\left[p_r(\vec{r}\,'_{\mathbf{\Sigma}})+
	(i\omega\tau_D\beta 
	a)(p_i(\vec{r}\,'_{\mathbf{\Sigma}})-R_S\,p_r(\vec{r}\,'_{\mathbf{\Sigma}})))\right]
	\,\widehat{u}\,dS'\nonumber
\end{eqnarray}
Since $p_r,p_i$ and $R_S$ are dimensionless functions determined solely by 
particles' shape, their surface integrals scale as $a^2$, and the dipole 
moment scales as $E_0 a^3\propto E_0 V$.  As the polarization is not 
necessarily collinear with $E_0\widehat{z}$, its final form is 
$(\overline{\overline{P}}_r+(a/\lambda)i\omega\tau_D~\overline{\overline{P}}_i 
)~\vec{E}_0~V$, just as in Proposition I.

Finally, we outline a proof of the existence of a maximum in the phase shift 
defined in Eq.~\ref{effective-def}. At low frequencies, dielectric enhancement 
dominates $P$ and $|\tan\theta(\omega)|\approx (a/\lambda)f\omega\tau_D|P_i|$ 
is a fast increasing function of $\omega$. At larger $\omega$, we show in 
online SI, by methods similar to those in this Appendix, that $P$ is real and 
has no enhancement, generating 
$\tan\theta(\omega)=(1+f\,P_r)\omega\tau_D\approx\omega\tau_D$.  It is then 
easy to find $\theta_L, \theta_H$, from the low and high frequency regimes, 
such that $|\tan\theta_L|>|\tan\theta_H|$. Thus $|\tan\theta(\omega)|$, 
initially increasing and then decreasing, must have a maximum. The details are 
given in online SI.

\newpage
\section{Supporting Information}
\subsection{Model Validity Conditions}\hspace{3pt}

In this section, we address the three conditions for our analysis in the main 
text: the limit on the driving field $E_0$, the validity for ignoring the 
hydrodynamic flow due to electro-osmosis and the limit on $\zeta$-potential.  
Our derivations of these conditions focus on the low frequency, universal 
enhancement regime discussed in the main text, but most can be generalized to 
a broader frequency range.

Our approach is based on self-consistency arguments, in which we show that, 
under the conditions given in the main text, the solution of our simplified, 
linear model is also the solution of the more complete model under discussion.  
Such arguments rely on the uniqueness of solutions of the complete model, which 
is reasonable physically, because the dielectric responses of random 
suspensions in electrolyte solution are not known to exhibit bistable 
behaviors. We will not, however, attempt mathematical proofs for such 
uniqueness here.
\subsubsection{Linearization Condition: {\small $eE_0a\ll k_BT\Rightarrow 
n_{\pm}\ll N_0$}}
We begin by deriving the condition for linearizing the charge conservation 
equations Eq.~\ref{charge_consv}. As we note in the main text, for symmetric 
ions, we have $n^{\textrm{total}}=n_++n_-=0$ throughout the electrolyte 
solution, so $|n_{\pm}|=(1/2)n^{\textrm{net}}$. Thus we only need to prove 
$|n^{\textrm{net}}\ll N_0|$ throughout the electrolyte solution.

Of course, $n^{\textrm{net}}$ is non-zero only within the EIDL. Furthermore, as 
shown in the main text, the it decays exponentially away from the 
solid-solution interfaces, so we only need to prove $n_0\ll N_0$, where $n_0$ 
is defined in Eq.~\ref{blocking_BC} in the main text. Thus we have:
\begin{equation}\label{eqn:n_0_N_0}
	\frac{1}{2}\frac{|n_0|}{N_0}
	=\frac{|E^{\bot}_{L}\beta\sigma_w|}{\omega N_0e}
	=\frac{|E^{\bot}_{L}\beta| e}{\omega\tau_D k_B T}\lambda^2
	=\frac{e|\psi_{\mathrm{EIDL}}|}{k_B T}\lambda^2|\beta|^2
	\approx\frac{e|\psi_{\mathrm{EIDL}}|}{k_B T}.
\end{equation}
Here we use the fact that $\sigma_w=\epsilon_0\epsilon_w'/\tau_D$, 
Eq.~\ref{blocking_BC} relating $n_0$ to $E^{\bot}$ and the definition of 
$\lambda$ in Eq.~\ref{debye}.  We also introduce the 
$\psi_{\mathrm{EIDL}}=\psi_L-\psi_S$, which is the potential drop within the 
EIDL and use the effective BC in Eq.~\ref{BC_eff}b to relate 
$\psi_{\mathrm{EIDL}}$ to $E^{\bot}_L$.

At low frequency, the potential on the surfaces just outside the EIDL 
$\psi_L\approx\psi^0_L$, which is the zeroth-order solution defined by the 
integral equation Eq.~\ref{eqn:zeroth-order}. As discussed in the main text, 
$\psi^0_L=E_0a p_r(\vec{r})$, and $p_r$ is a dimensionless function determined 
solely by the shape of the particle via Eq.~\ref{eqn:zeroth-order}.

As long as the particle is smooth, convex and has an aspect ratio not too far 
from one (\emph{i.e.} not shaped like a disk or needle), we expect the integral 
equation gives $p_r(\vec{r})$ of order one throughout the surface 
$\mathbf{\Sigma}$ of the particle, as there is no singularity or 
\emph{geometric} parameters that are much bigger or smaller than one. Thus we 
expect the potential on the surface to be of the order $\psi_L\sim E_0a$.

A comparison with the Philips integrals for ordinary dielectrics~\cite{Philips} 
shows that, at this limit, the external potential $\psi^0_L$ is approximately 
that outside a particle with $\epsilon'=0$.  If the particle is smooth and of 
aspect ratio not far from one, we expect the electric field around the particle 
to be comparable to $E_0$ (and has no component normal to the particle 
surface).  As we generally set the origin of the potential to be inside the 
particle, on its surface the potential $\psi_L$ is of the order $E_0a$.

For the internal potential, we begin by noting that $\psi_S'$, as given by
Eq.~\ref{equ:R-self-consistent}, is clearly proportional to $E_0$ as $\psi_0$ 
does. Moreover, a rescaling of particle size $a\to\alpha a$ will result in 
$\psi_S'\to\alpha\psi_S$. Thus $\psi_S'\propto E_0a$, with the proportionality 
constant determined entirely by the shape of the particle, with no extra length 
scale like $\lambda$ in Eq.~\ref{eqn:perturbation}. Thus we expect for 
particles with smooth, convex surface and aspect ratios not far from one, 
$\psi_S'\sim E_0a$, so $|\psi_S|=|i\omega\tau_D\beta a\epsilon_L\psi_S'|\ll 
E_0a$ because we are at such a low frequency  that
$\omega\tau_D\ll(\lambda/a)(\sigma_s/\sigma_w)\approx|1/(\beta a\epsilon_L)|$.


So we have across the surface $\mathbf{\Sigma}$,
$\psi_{\mathrm{EIDL}}=\psi_L-\psi_S\sim E_0a$. Thus by Eq.~\ref{eqn:n_0_N_0} 
when $eE_0a\ll k_B T$, we have $|n_\pm|\ll N_0$ and the linearization in the 
main text is valid.
\subsubsection{The Insignificance of the Electrophoresis Flow}
For electrokinetic flow, we continue the planar approximation by which we 
derive the effective BCs in the EIDL and use the classic Helmholtz-Smoluchowski 
formula for planes~\cite{Levich}:
\begin{equation}\label{eqn:flow-speed}
	|v^{\parallel}_{\mathrm{max}}|=
	\frac{\epsilon_0\epsilon_w'}{\eta}E^{\parallel}|\psi_{\mathrm{slip}}-\psi(\xi=0)|
	\le\frac{\epsilon_0\epsilon_w'}{\eta}E^{\parallel}|\psi_{\mathrm{EIDL}}|.
\end{equation}
Here $\eta$ is the dynamic viscosity of water and 
$v^{\parallel}_{\mathrm{max}}$ is the maximum relative velocity the solution 
reaches. $E^{\parallel}$ is the electric field tangential to the particle 
surface, and it varies relatively slowly within the EIDL compared with the 
normal field.  $\psi_{\mathrm{slip}}$ is the potential at the slipping plane in 
the EIDL.  Since in our EIDL, the potential drop monotonically decreases, the 
maximum amount of potential drop within it is $\psi_{\mathrm{EIDL}}$, as 
defined above.  The maximum ionic current associated with the electro-osmosis 
flow is thus 
$j^{\mathrm{drag}}_{\mathrm{max}}=v^{\parallel}_{\mathrm{max}}N_0$.

Let us compare this with the conductive current due to the electric field 
$E^{\parallel}$, 
$j^{\mathrm{cond}}=E^{\parallel}\sigma_w/e=E^{\parallel}\epsilon_0\epsilon_w'/e\tau_D$, 
which is but one component of the ionic current considered in 
Eq.~\ref{charge_consv}. The ratio between the two is:
\begin{equation}\label{eqn:flow-current-comparison}
	\frac{|j^{\mathrm{drag}}|}{|j^{\mathrm{cond}}|}\le
	\frac{e|\psi_{\mathrm{EIDL}}|N_0}{\eta/\tau_D}\ll
	\frac{k_B T N_0}{\eta/\tau_D}\approx0.134.
\end{equation}
The last ratio turns out to be independent of $N_0$ after taking into account 
of the $N_0$ dependence of $\tau_D$.
In the last step we used the following material parameters: $D=2\times10^{-9} 
m^2/s, T=300K, \eta=8.9\times10^{-4}Pa\cdot s$. Thus the electrokinetic flow is 
much smaller than the conductive flow in the planar, linearized EIDL 
approximation of our model.
\subsubsection{$\zeta$-potential is unimportant for {\small $e\psi_{\zeta}\ll 
k_B T$}}
We show in this subsection that when the $\zeta$-potential $e\psi_{\zeta}\ll 
k_BT$ the equations of motion for the time-dependent physics reduces to that of 
Eq.~\ref{eq:motion} and produces the same enhancement as in the main text.

We begin by analyzing the physics under finite $\zeta$-potential in the static 
limit without the external drive $E_0\widehat{z}\exp(i\omega t)$. We still 
assume the minimal radius of curvature of all interfaces are much larger than 
the Debye length $\lambda$, so the interfaces are considered infinite planes 
and the problem reduced to that of 1D. The system is assumed to be in local 
equilibrium, so that the density currents $\vec{j}_{\pm}$, as defined in 
Eq.~\ref{charge_consv}, are everywhere zero and the ion densities follow the 
Boltzmann distribution. The equations of motion are then:
\begin{equation}\label{eqn:static-motion}
	N_{\pm}=N_0e^{\mp\frac{e\psi}{k_B T}},\,\,\,
	\nabla^2\psi_0\approx 
\frac{d^2\psi}{d\xi^2}=-\frac{e(N_+-N_-)}{\epsilon_0 \epsilon_w'}.
\end{equation}
Here $N_0$ is the ion density in the charge-neutral liquid outside the double 
layer.  The solution of this equation is well-known~\cite{Chew}.
\begin{equation}\label{eqn:static-solution}
	\frac{e\psi_0(\xi)}{k_B T}=2\ln\left(\frac{1+t e^{-\xi/\lambda}}{1-t 
	e^{-\xi/\lambda}}\right),\,\,\,t=\tanh\left(\frac{e\psi_{\zeta}}{4k_B 
	T}\right).
\end{equation}
When the $\zeta$-potential is small, $e\psi_{\zeta}\ll k_BT$, $t\ll1$:
\begin{equation}\label{eqn:static-low-psi}
	\frac{e\psi_0(\xi)}{k_B T}\approx e^{-\xi/\lambda},\,\,\, 
	N_{\pm}(\xi)=N^{\pm}_0 e^{-\xi/\lambda},\,\,\,
	N^{\pm}_0=N_0 e^{\mp e\psi_{\zeta}/k_B T}\approx N_0.
\end{equation}

Now we turn on the external field $E_0\widehat{z}\exp(i\omega t)$. We write the 
time-dependent solution as $N^t_{\pm}=N_{\pm}+n_\pm$ and $\psi^t=\psi_0+\phi$.  
We assume here \emph{only} that $n_{\pm}\ll N_{\pm}\approx N_0$, which is the 
same as the linearization condition discussed above. {\em We do not assume the 
potential change $\phi$ is small compared with $\psi$}.

Using Eq.~\ref{eqn:static-motion}, it is easy to see that Poisson equation 
preserves its form for $n_{\pm}$ and $\phi$:
\begin{equation}\label{eqn:poisson-finite-zeta}
\nabla^2\phi= -\frac{e(n_+-n_-)}{\epsilon_w'\epsilon_0}.
\end{equation}
The ionic density currents are more complicated because of the cross terms, but 
using the fact that the static ionic currents are zero anywhere, and $n_\pm\ll 
N_0\approx N_\pm$, we have:
\begin{equation}\label{eqn:two-facts}
	\vec{j}_{\pm}=
	-D\left(\overrightarrow{\nabla} N_\pm \pm
	\frac{e N_\pm}{k_BT}\overrightarrow{\nabla}\psi\right)=0,\,\,
\left|\frac{e n_\pm}{k_BT}\overrightarrow{\nabla}\phi\right|\ll
\left|\frac{e N_\pm}{k_BT}\overrightarrow{\nabla}\phi\right|,
\end{equation}
and the time-dependent current densities simplify:
\begin{eqnarray}\label{eqn:current-finite-zeta}
	\vec{j}_{\pm}^{\, t}&=&
	-D\left(\overrightarrow{\nabla} N^t_\pm \pm
	\frac{e N^t_\pm}{k_BT}\overrightarrow{\nabla}\psi^t\right)\nonumber\\
	&\approx&-D\left(\overrightarrow{\nabla}n_\pm\pm\left(
\frac{e n_\pm}{k_BT}\overrightarrow{\nabla}\psi
+\frac{e N_0}{k_BT}\overrightarrow{\nabla}\phi \right)\right)
\end{eqnarray}
Now assume $\phi$ is the potential inside the double layer that, like those of 
the solution in the main text, varies spatially as 
$\psi_{\mathrm{EIDL}}\exp(-\xi/\lambda)$.
Because when the $\zeta$-potential is weak $e\psi_{\zeta}\ll k_B T$, the 
potential $\psi$ inside the double layer is $\psi_{\zeta}\exp(-\xi/\lambda)$ 
(Eq.~\ref{eqn:static-low-psi}), the two conductive current terms in 
Eq.~\ref{eqn:current-finite-zeta} can be easily compared:
\begin{equation}\label{eqn:finite-zeta-comp}
	\frac
	{e n_\pm/k_BT|\overrightarrow{\nabla}\psi|}
	{e N_0/k_BT|\overrightarrow{\nabla}\phi|}
	=
	\frac
	{e \psi_{\zeta}/k_BT\,n_\pm}
	{e \psi_{\mathrm{EIDL}}/k_BT\,N_0}
	=\frac{e\psi_{\zeta}}{k_BT}\frac{n_\pm/N_0}{e\psi_{\mathrm{EIDL}}/k_BT}
\end{equation}
According to Eq.~\ref{eqn:n_0_N_0}, the second factor is approximately two, and 
by assumption the first is much smaller than one. Thus, the time dependent 
current densities have forms completely analogous to Eq.~\ref{charge_consv} in 
the main text:
\begin{equation}\label{eqn:current-final}
	\vec{j}^{\,t}_{\pm}\approx
	-D\left(\overrightarrow{\nabla}n_\pm\pm
	\frac{e N_0}{k_BT}\overrightarrow{\nabla}\phi\right).
\end{equation}
Combining Poisson equation Eq.~\ref{eqn:poisson-finite-zeta} and currents
Eq.~\ref{eqn:current-final}, one see that when the $\zeta$-potential is small, 
$\phi$ and $n_\pm$ follows exactly the equations of motion in the main text.

Furthermore, the equations for $\psi,\phi$ and for $N_{\pm},n_\pm$ completely 
decouple, and the physics is determined by the linear superposition of the two.  
The static solution $\psi_0$ obviously only makes a real, frequency-independent 
contribution to the total polarization, so when $e\psi_{\zeta}\ll k_B T$, the 
$\zeta$-potential does not change the dielectric enhancement described in the 
main text.
\subsection{Asymmetric Ions}\hspace{3pt}

In the main text, we have made the simplifying assumption that the ions are 
symmetric, with cation and anion having not only the same charge, but also the 
same diffusion coefficient. The second assumption is clearly unrealistic. In 
this section we show that relaxing these two assumptions do not change the 
physical picture in the main text.

Assume the diffusion coefficient of the cations and anions are $D_+,D_-$, their 
charges are $q_+,-q_-$, and their density outside the EIDL are $N_+,N_-$. Also 
assume that outside EIDL the liquid is charge neutral: $N_-q_-=N_+q_+$. The 
total and net particle densities no longer separate naturally, so 
Eq.~\ref{eq:net_motion} no longer hold. Instead, we have
{\setlength\arraycolsep{0.5pt}
\begin{eqnarray}\label{eqn:asym-motion}
	\nabla^2n_+&=&(\alpha_++\frac{1}{\lambda_+^2})n_+-\frac{1}{\lambda_-^2}n_-,\,\,\,\,\,\,\alpha_{\pm}=\frac{i\omega}{D_\pm},
	\nonumber\\
	\nabla^2n_-&=&-\frac{1}{\lambda_+^2}n_++(\alpha_-+\frac{1}{\lambda_-^2})n_-,\,\,\,\,
	\lambda_{\pm}^2 = \frac{k_BT\,\epsilon_0 \epsilon_w'}{N_\pm q_{\pm}^2}.
\end{eqnarray}}
On the other hand, the boundary condition Eq.~\ref{eq:bc} are still best 
expressed in terms of $n^{\mathrm{total}}=n_++n_-$ and 
$n^{\mathrm{net}}=n_+-n_-$:
\begin{eqnarray}\label{eqn:asym-BC}
	\widehat{u}\cdot\overrightarrow{\nabla}n^{\mathrm{net}}&=&-\frac{2\epsilon_0\epsilon_w'}
	{\lambda_{\pm}^2q_{\pm}}\widehat{u}\cdot\overrightarrow{\nabla}\psi\nonumber\\
	\widehat{u}\cdot\overrightarrow{\nabla}n^{\,\mathrm{total}}&=&0
\end{eqnarray}
The obvious way forward is to make linear combinations of $n_{\pm}$ to make 
Eq.~\ref{eqn:asym-motion} diagonal. To keep algebras and notations manageable 
and the physics transparent, here we focus on the case where the ions have the 
same charge $q_+=q_-=q$ (hence $N_+=N-=N_0, 
\lambda_+^2=\lambda_-^2=2\lambda^2$) but different diffusion coefficients.  
Under these transformations:
{\setlength\arraycolsep{0.5pt}
\begin{eqnarray}\label{eqn:asym-transformation}
	n^{\mathrm{net}}&=&-\lambda^2\alpha_2~\rho_1+\left(1-\lambda^2\alpha_1\right)~\rho_2,\,\,\,\,\alpha_1=\frac{1}{2}\left(\alpha_++\alpha_-\right)
	\nonumber\\
	n^{\mathrm{total}}&=&\left(1+\lambda^2\alpha_1\right)~\rho_1+\lambda^2\alpha_2~\rho_2,\,\,\,\,\alpha_2=\frac{1}{2}\left(\alpha_+-\alpha_-\right),
\end{eqnarray}}
the equations of motion are diagonal:
\begin{eqnarray}\label{eqn:asym-transformed}
	\nabla^2\rho_1&=&\left(\frac{1}{\lambda^2}+\alpha_1\right)\rho_1,\nonumber\\
	\nabla^2\rho_2&=&\alpha_1\rho_2,\nonumber\\
	\nabla^2\psi&=&-\frac{q}{\epsilon_0\epsilon_w'}n^{\mathrm{net}}.
\end{eqnarray}
Combining Eq.~\ref{eqn:asym-BC}--\ref{eqn:asym-transformed}, and following the 
derivation of the EIDL approximation in the main text, one can recover the 
following effective BCs analogous to Eq.~\ref{BC_eff}:
\begin{eqnarray}\label{eqn:asym-eff-BC}
	E^{\bot}_{S}&=&c\left(1+\frac{1}{i\omega\widetilde{\tau}}\right)E^{\bot}_{L},\nonumber
	\\
	\psi_{S}&=&\psi_{L}+\frac{1}{i\omega\widetilde{\tau}\widetilde{\beta}} 
	E^{\bot}_{L},
\end{eqnarray}
where we redefine the parameters $\tau_D$ and $\beta$ as following
\begin{equation}\label{eqn:asym-def}
	\widetilde{\tau}
	=\frac{\lambda^2}{\widetilde{D}},\,\,\,\,\,\,
	\widetilde{\beta}^2=\frac{1}{\lambda^2}(1+i\omega\overline{\tau}),\,\,\,\,\,\,
	\overline{\tau}=\frac{\lambda^2}{\overline{D}},
\end{equation}
and $\overline{D}$ and $\widetilde{D}$ are the harmonic and arithmetic mean of 
the two diffusion constants:
\begin{equation}\label{eqn:diffusion-const}
	\overline{D}=\frac{2}{1/D_1+1/D_2},\,\,\,\,\,\,
	\widetilde{D}=\frac{D_1+D_2}{2}.
\end{equation}
The case of ions with unequal charges proceed similarly, albeit with much more 
complicated redefinitions of $\tau$ and $\beta$. But the form of the effective 
BCs remain those of Eqs.~\ref{eqn:asym-eff-BC}.

Of course, with boundary conditions like Eqs.~\ref{eqn:asym-eff-BC}, reasoning 
about dielectric enhancement identical to those in the main text can be carried 
out, with identical conclusions and merely redefined parameters.

\subsection{Derivation of the Philips Integral Formulation}\hspace{3pt}

To derive the Philips integral equation formulation for our BCs 
Eq.~\ref{BC_eff_low_freq}, we use the fact that potentials on both side of the 
EIDL obey Laplace equation, and make use of the following form of the Stokes' 
theorem for a harmonic function to compute its value anywhere in a domain 
$\mathbf{\Omega}$ from its value on it boundary $\vec{r}\,'_{\mathbf{\Sigma}}$:
{\setlength\arraycolsep{2pt}
\begin{eqnarray}\label{eq:stokes}
	&&\int_{\mathbf{\Sigma}}\left(A(\vec{r}\,'_{\mathbf{\Sigma}})\widehat{u}\cdot\overrightarrow{\nabla}_{\vec{r}\,'_{\mathbf{\Sigma}}}B(\vec{r},\vec{r}\,'_{\mathbf{\Sigma}})-\widehat{u}\cdot\overrightarrow{\nabla}_{\vec{r}\,'_{\mathbf{\Sigma}}}A(\vec{r}\,'_{\mathbf{\Sigma}})B(\vec{r},\vec{r}\,'_{\mathbf{\Sigma}})\right)dS'\nonumber\\
	&=&\left\{
	\begin{array}{rl}
		A(\vec{r})&\forall \vec{r}\in\mathbf{\Omega},\\
		0&\forall \vec{r}\notin\mathbf{\Omega}.
	\end{array}
	\right.
\end{eqnarray}}
Here $B=-\frac{1}{4\pi}\frac{1}{|\vec{r}-\vec{r}\,'|}$ is the Green's function 
of the Laplace operator. $A(\vec{r})$ is a harmonic function that is non-zero 
in a simply-connected domain $\mathbf{\Omega}$ and zero elsewhere, and the 
\emph{surface} integral is carried out on $\vec{r}\,'_{\mathbf{\Sigma}}$ which 
lies on the boundary $\mathbf{\Sigma}=\partial\mathbf{\Omega}$. Any variable 
with subscript $_\mathbf{\Sigma}$ lies on $\mathbf{\Sigma}$.  $\widehat{u}$ is 
a unit normal vector pointing \emph{out of} the domain $\mathbf{\Omega}$.  In 
this section the boundary $\mathbf{\Sigma}$ is set to the EIDL, which is 
considered to be infinitely thin compared with the solid and neutral liquid 
domains.

Let function $\psi_S$ coincide with the potential in the solid and be zero in 
the neutral liquid outside the EIDL, and function $\psi_L$ coincide with the 
potential in the neutral liquid and be zero within the solid. To apply
Eq.~\ref{eq:stokes} to the unbounded domain in the neutral liquid, the function 
$A(\vec{r})$ need to vanish quicker than $1/r^2$ at infinity. Thus we need to 
subtract from $\psi_L$ (and $\psi_S$ for symmetry reason) the ``incident'' 
potential $\psi_0=-E_0z$ due to the external drive.  Set 
$A=\psi_{S}/\epsilon_L-\psi_0$ and again $A=\psi_L-\psi_0$, then
Eq.\,\ref{eq:stokes} for a point $\vec{r}$ lying \emph{inside the solid} become
{\setlength\arraycolsep{0.5pt}
\begin{eqnarray}\label{eqn:philips}
	&&\frac{\psi_{S}(\vec{r})}{\widetilde{\epsilon}_S}-\psi_0(\vec{r})=
\frac{1}{4\pi}\oiint\limits_{\mathbf{\Sigma}}
\Bigg\{\left[\frac{\psi_{S}(\vec{r}\,'_{\mathbf{\Sigma}})}{\widetilde{\epsilon}_{S}}-\psi_0(\vec{r}\,'_{\mathbf{\Sigma}})\right]\,
	\widehat{u}\cdot\overrightarrow{\nabla}_{\vec{r}\,'_{\mathbf{\Sigma}}}\frac{1}{|\vec{r}-\vec{r}\,'_{\mathbf{\Sigma}}|}\nonumber\\
	&&-
	\left[\widehat{u}\cdot\overrightarrow{\nabla}_{\vec{r}\,'_{\mathbf{\Sigma}}}\psi_{S}(\vec{r}\,'_{\mathbf{\Sigma}})/\widetilde{\epsilon}_{S}-\widehat{u}\cdot\overrightarrow{\nabla}_{\vec{r}\,'_{\mathbf{\Sigma}}}\psi_0(\vec{r}\,'_{\mathbf{\Sigma}})\right]\,
\frac{1}{|\vec{r}-\vec{r}\,'_{\mathbf{\Sigma}}|}
\Bigg\}dS',\\
	&&0=
\frac{1}{4\pi}\oiint\limits_{\mathbf{\Sigma}}
\Bigg\{\left[\psi_{L}(\vec{r}\,'_{\mathbf{\Sigma}})-\psi_0(\vec{r}\,'_{\mathbf{\Sigma}})\right]\,
	\widehat{u}\cdot\overrightarrow{\nabla}_{\vec{r}\,'_{\mathbf{\Sigma}}}\frac{1}{|\vec{r}-\vec{r}\,'_{\mathbf{\Sigma}}|}\nonumber\\
	&&-
\left[\widehat{u}\cdot\overrightarrow{\nabla}_{\vec{r}\,'_{\mathbf{\Sigma}}}\psi_{L}(\vec{r}\,'_{\mathbf{\Sigma}})-\widehat{u}\cdot\overrightarrow{\nabla}_{\vec{r}\,'_{\mathbf{\Sigma}}}\psi_0(\vec{r}\,'_{\mathbf{\Sigma}})\right]\,
\frac{1}{|\vec{r}-\vec{r}\,'_{\mathbf{\Sigma}}|}
\Bigg\}dS',
\end{eqnarray}}
Subtract the two equations. Use BC Eq.~\ref{BC_eff}a $
\widehat{u}\cdot\overrightarrow{\nabla}\psi_S/\widetilde{\epsilon}_L=\widehat{u}\cdot\overrightarrow{\nabla}\psi_L$
to cancel the second terms inside the integral. Then use BC  Eq.~\ref{BC_eff}b 
to eliminate $\psi_L$, with $\widehat{u}\cdot\overrightarrow{\nabla}\psi_L$ 
being substituted by $\widehat{u}\cdot\overrightarrow{\nabla}\psi_S$ via 
Eq.~\ref{BC_eff}a. One then arrives at Eq.~\ref{integral_eq} for $\psi_S$, 
expressing the value of $\psi_S$ inside the solid by its value on the 
boundary.  The same can be done for a point $\vec{r}$ lying \emph{in the 
charge neutral liquid}, which gives the equivalent expression for 
$\psi_L(\vec{r})$.

Eq.~\ref{eqn:potential-in-bulk} and Eq.~\ref{effective_potential} are identical 
to that of the simple dielectric. The only difference is the extra term due to 
the dipole moment in the EIDL $1/(i\omega\tau_D\beta)\,
\widehat{u}\cdot\overrightarrow{\nabla}\psi_{\mathsf{\Omega}}$, which produces 
the dielectric enhancement.

Finally, the derivation of Eq.~\ref{integral_eq} from 
Eq.~\ref{eqn:potential-in-bulk}, by taking the $\vec{r}$ to the surfaces, is 
standard. The only subtlety is the singularity of the kernel
$\overrightarrow{\nabla}(1/|\vec{r}-\vec{r}\,'|)$, which is equivalent to that 
of a dipole layer on the interface $\mathbf{\Sigma}$. For a smooth surface, 
this singularity contribute an extra term equal to $\pm1/2$ times the value of 
the function multiplying the kernel at point $\vec{r}$~\cite{Philips,Bladel}, 
depending the direction of the normal vector $\widehat{u}$.
\subsection{Scale Invariant Maximum in the Impedance Phase Shift}\hspace{3pt}

We first note that condition $\sigma_s\gg\sigma_w$ is not very restrictive in 
real world applications, because even rather strong electrolyte like sea water 
and human blood has conductivity of the order a few $S/m$. Not only any metal, 
but common engineering materials like amorphous carbon and moderately doped 
semiconductor have several orders of magnitude higher conductivities than the 
electrolyte solution.

Secondly, when $\sigma_s\gg\sigma_w$, the condition 
Eq.\,\ref{enhancement_condition_material} allows a much wider frequency range 
than Eq.\,\ref{enhancement_condition_geometry}. All our discussions, including 
the ``high-frequency'' regime below, happen under the range of 
Eq.~\ref{enhancement_condition_material} (\emph{c.f.} 
Eq.~\ref{eqn:high-frequency-condition}).  Under this condition, the expansion 
Eq.~\ref{epsilon_eff_out} of the effective potential for $\psi_L$ is valid, so 
there is always the cancellation of $\widetilde{\epsilon}_L$ in 
Eq.~\ref{eqn:perturbation}. This means that the material property of the 
particle $\sigma_s/\sigma_w$ drops out of the integral equation for $\psi_L$ so 
anything determined by the dipole moment $P$ will be material independent.  
This shows that the frequency-dependent phase shift angles $\tan\theta(\omega)$ 
is independent of $\sigma_s/\sigma_w$ at the frequencies where 
Eq.~\ref{enhancement_condition_material} hold.

At low frequency when conditions in 
Eqs.\,\ref{enhancement_condition_material},~\ref{enhancement_condition_geometry}
are satisfied, the enhanced part of polarization, due to the imaginary part of 
the dipole moment $P_i$, dominates, so
\begin{equation}\label{eqn:phase_low}
|\tan\theta(\omega)|\approx (a/\lambda)\omega\tau_D|P_i|\,\,f
\end{equation}
increase monotonically with $\omega$. Thus, if we can show that at a larger 
frequency, $|\tan\theta(\omega)|$ decrease in value, there must be a maximum in 
a frequency between.

At higher frequency, consider the following conditions:
\begin{equation}\label{eqn:max-condition}
	\omega\tau_D\gg\frac{1}{R'}\frac{\lambda}{a}\,\,\mathbf{[a]},\,\,\,
	\omega\tau_D\ll\frac{\sigma_s}{\sigma_w}\frac{\epsilon_w}{\epsilon_s},\,
	1\,\,\mathbf{[b]},\,\,\,
	R'=\min_{\vec{r}\mathbf{\in\Sigma}}\,|R_S(\vec{r})|.
\end{equation}
$R_S$ is defined in Eq.~\ref{eq:R-def} in the main text. When both conditions 
in  Eq.~\ref{eqn:max-condition}b hold, 
$\epsilon_L\approx\sigma_w/\sigma_s\ll1$.  Under condition 
Eq.~\ref{eqn:max-condition}a, we also have, across the interface, 
$|\widehat{u}\cdot\overrightarrow{\nabla}\psi_{S}/(i\omega\tau_D\beta)|\ll|\psi_S|$.  
Combining the two, we have
$|\psi^{\mathrm{eff}}_S|\ll|\psi_S|$. Upon a simple rescaling 
$\psi_S'=\psi_S/\widetilde{\epsilon}_L$, which does not affect $R_S$ (see 
Eq.~\ref{eq:R-def}), the integral equation Eq.~\ref{integral_eq} for $\psi_S'$ 
again reduces to a very simple form:
\begin{equation}\label{eqn:psi_S_high_frequency}
	\psi_0(\vec{r})=
	\frac{1}{2}
	\psi_S'(\vec{r})
	+\frac{1}{4\pi}\oiint\limits_{\mathbf{\Sigma}}
	\psi_S'(\vec{r}\,')
	\widehat{u}\cdot\overrightarrow{\nabla}_{\vec{r}\,'}\frac{1}{|\vec{r}-\vec{r}\,'|}dS'
\end{equation}
It is easy to see that under these conditions $R_S$, and by extension $R'$, are 
again determined solely by a particle's shape. As argued in the main text, for 
smooth, convex shapes whose aspect ratios are not too far from one, we expect 
$R'$ to be of order one.

Now turn to $\psi_L$ under this limit. Since $\widetilde{\epsilon}_L\ll1$, we 
can find a frequency that satisfies
\begin{equation}\label{eqn:high-frequency-condition}
1\ll|i\omega\tau_D\beta a R_S|\ll 1/\widetilde{\epsilon_L}
\end{equation}
across the interface (except, again, those point where $R_S$ diverges, which as 
we have argued in the main text can be ignored if we choose $|\omega\tau_D\beta 
a|$ sufficiently far from the upper limit $\epsilon_L$). At such a frequency, 
the form of the effective potential Eq.~\ref{epsilon_eff_out} remains true, so 
we still have:
{\setlength\arraycolsep{0.0pt}
\begin{eqnarray}\label{eqn:psi_L_high_frequency}
	\psi_0(\vec{r})&=&
	\frac{1}{2}\left(1+
	i\omega\tau_D\beta a R_S\right)
	\psi_L(\vec{r})\\
	&-&\frac{1}{4\pi}\oiint\limits_{\mathbf{\Sigma}}
	\left(1-
	i\omega\tau_D\beta a R_S\right)
	\psi_L(\vec{r}\,')
	\widehat{u}\cdot\overrightarrow{\nabla}_{\vec{r}\,'}\frac{1}{|\vec{r}-\vec{r}\,'|}dS',\nonumber
\end{eqnarray}}
But now, by Eq.~\ref{eqn:max-condition}a, the effective potential term is 
dominant. Thus, after introducing a rescaling 
$\psi_L'=\psi_L(i\omega\tau_D\beta a)$, we can again transform the integral 
equation into a very simple form:
{\setlength\arraycolsep{0.0pt}
\begin{equation}\label{eqn:psi_L_high_frequency_simple}
	\psi_0(\vec{r})=
	\frac{1}{2}
	R_S(\vec{r})\psi_L'(\vec{r})
	+\frac{1}{4\pi}\oiint\limits_{\mathbf{\Sigma}}
	R_S(\vec{r}\,')\psi_L'(\vec{r}\,')
	\widehat{u}\cdot\overrightarrow{\nabla}_{\vec{r}\,'}\frac{1}{|\vec{r}-\vec{r}\,'|}dS'.
\end{equation}}
By same scaling arguments in the main text, we expect $\psi'(\vec{r})=E_0a\, 
p(\vec{r})$, where $p$ is a dimensionless function determined solely by 
particle's shape.

At high frequency described in Eq.~\ref{eqn:max-condition}a, the dipole 
integral analogous to Eq.~\ref{eqn:dipole} in our case is:
\begin{eqnarray}\label{eqn:dipole-high-frequency}
	&&\psi_L(\vec{r})=
	\psi_0(\vec{r})
	+\frac{1}{4\pi}\oiint\limits_{\mathbf{\Sigma}}
	\psi_{L}\left(1-i\omega\tau_D\beta a R_S\right)\,
	\widehat{u}\cdot\frac{\vec{r}}{|\vec{r}|^3} dS'\nonumber\\
	&&\approx
	\psi_0(\vec{r})-
	\frac{1}{4\pi}\frac{\vec{r}}{|\vec{r}|^3}\cdot
	\oiint\limits_{\mathbf{\Sigma}}\,\psi_L'(\vec{r}\,')\,R_S(\vec{r}\,')\widehat{u}\,dS'.
\end{eqnarray}
By the same argument in the main text, the polarization at this limit will 
scale as $E_0a^3=E_0 V$. And it will obviously be purely real, as $\psi'_L$ and 
$p(\vec{r})$ are purely real. 

More important, any additional length scale, such as $\lambda$ in the imaginary 
part of polarization in the enhancement regime in Eq.~\ref{eqn:P_scaling}, is 
absent here.  Indeed, if the particle's surface is smooth, convex and its 
aspect ratios not too far from one, there is no additional very small or very 
large \emph{geometric} parameters in Eq.~\ref{eqn:psi_L_high_frequency_simple}, 
so we expect the polarization in this limit, upon orientation averaging, to be 
of the form $E_0 V\,P$, where $P$ is a real, dimensionless number \emph{of 
order one}, independent of both frequency and material parameters.
Thus, in this limit, 
$\epsilon'_{\mathrm{eff}}=\epsilon_w'(1+f\,P)\approx\epsilon_w'$ and
\begin{equation}\label{eqn:phase-high}
\tan\theta(\omega)\approx\omega\tau_D.
\end{equation}

Because of the large factor $(a/\lambda)\,f$ in Eq.~\ref{eqn:phase_low} and its 
absence in Eq.~\ref{eqn:phase-high}, it should not be difficult to find a 
frequency $\omega_{L}$ in the regime 
Eqs.\,\ref{enhancement_condition_material},\ref{enhancement_condition_geometry}
and a frequency $\omega_H$ in the regime 
Eqs.\,\ref{eqn:high-frequency-condition} such that 
$\tan\theta(\omega_L)>\tan\theta(\omega_H)$. This will prove that the function 
$\tan\theta(\omega)$ decreases after initially increasing monotonically and 
thus have a maximum between $\omega_L$ and $\omega_H$.

To be more precise, assume we have
\begin{equation}\label{eqn:frequency-factor-definition}
	\omega_{L,H}\,\tau_D=\frac{\lambda}{a}A_{L,H},\,\,\, A_L\ll 
	\frac{1}{R},\,\,\,\frac{1}{R'}\ll A_H\ll \frac{\sigma_s}{\sigma_w}.
\end{equation}
Here $R$ and $R'$ are defined in Eq.~\ref{eq:R-def} and 
Eq.~\ref{eqn:max-condition}.  Then we have
\begin{equation}
	\frac{|\tan\theta(\omega_L)|}{|\tan\theta(\omega_H)|}
	\approx\frac{(a/\lambda)f|P_i|\omega_L\tau_D}{\omega_H\tau_D}
	=f |P_i|\frac{a}{\lambda}\frac{A_L}{A_H}.
\end{equation}
The factors in front of $A_L/A_H$ is essentially the size of dielectric 
enhancement factor at low frequency, which, as we mention in the abstract, can 
easily amount to $10^4$ even for a very dilute electrolyte solution. If we can 
assume that, for smooth, convex particles with aspect ratios not too far from 
one, because there is no additional \emph{geometric} small or large parameters 
far from one, both $R$ and $R'$ are not far from one, it should thus be not 
difficult to find a pair $A_L,A_H$ that satisfies both 
Eq.~\ref{eqn:frequency-factor-definition} and the condition 
$A_H/A_L<f|P_i|(a/\lambda)$, which, as discussed above, guarantee a maximum for 
the phase shift $|\tan\theta(\omega)|$ between $\omega_L$ and $\omega_H$. As we 
note above, this maximum will be independent of the material parameter 
$\sigma_s/\sigma_w$.

Finally, we want to show that the three approximations we use to derive the 
equations of motion and the effective BCs Eqs.~\ref{BC_eff} are still valid in 
the high frequency regime 
Eq.~\ref{eqn:max-condition},~\ref{eqn:high-frequency-condition},~\ref{eqn:frequency-factor-definition}.  
According the first section of this Online Support Information, the key is to 
prove that the potential drop across the EIDL $\psi_{\mathrm{EIDL}}$ is 
comparable or smaller than $E_0a$.

For $\psi_L$ on the charge neutral solution side, we have 
$\psi_L'=\psi_L(i\omega\tau_D\beta a)$ obeys 
Eq.~\ref{eqn:psi_L_high_frequency_simple}. As we argue above, $\psi_L'=E_0a\, 
p(\vec{r})$ where $p(\vec{r})$ is a dimensionless function determined solely by 
the shape of the particle. If the particle is smooth, convex and has aspect 
ratio not far from one, we expect $p(\vec{r})$ to be of order one. Thus, 
$\psi_L$ is of order $|E_0a|/|\omega\tau_D\beta a|$ on the interface. Assuming 
$R'$, as defined in Eq.~\ref{eqn:max-condition}, is of order one, as we have 
argued above from Eq.~\ref{eqn:psi_S_high_frequency}, 
Eq.~\ref{eqn:max-condition}a implies $|\omega\tau_D\beta a|\gg1$ and thus 
$\psi_L\ll E_0a$.

On the other hand, similar argument suggest that for the internal potential, 
$\psi_S'$ defined in Eq.~\ref{eqn:psi_S_high_frequency} is also of order $E_0a$ 
across the interface, meaning the potential on the solid side of the EIDL 
$\psi_S=\widetilde{\epsilon}_L\psi_S'\ll E_0a$, given that 
$\widetilde{\epsilon}_L=\sigma_w/\sigma_s\ll1$.

Combining the estimates for $\psi_L$ and $\psi_S$ we have 
$|\psi_{\mathrm{EIDL}}|=|\psi_L-\psi_S|\ll E_0 a$ and thus all three 
assumptions for deriving our linearized equation of motions hold at the 
frequency range of $\omega_H$ in Eq.~\ref{eqn:frequency-factor-definition} 
above.

\end{document}